\documentclass[aps,prb,a4paper,10pt,twocolumn,showpacs,floatfix,superscriptaddress,preprintnumbers,longbibliography]{revtex4-1}
\usepackage{float}
\usepackage{dcolumn,graphicx,color,booktabs,microtype,afterpage}
\usepackage{amssymb}
\usepackage{amsmath}
\usepackage[charter,greekuppercase=italicized]{mathdesign}
\usepackage{sidecap}
\usepackage[mathlines]{lineno}

\graphicspath{{./}{figure/}}
\renewcommand{\tablename}{Table}
\makeatletter\renewcommand{\fnum@figure}[1]{\figurename~\thefigure.~}\makeatother
\makeatletter\renewcommand{\fnum@table}[1]{\tablename~\thetable.}\makeatother

\newcount\hh \newcount\mm
\hh=\time \divide\hh by 60
\mm=\hh \multiply\mm by 60 \mm=-\mm
\advance\mm by \time
\def\now{\number\hh:\ifnum\mm<10{}0\fi\number\mm}

\usepackage[colorlinks,plainpages=false,linkcolor=blue,urlcolor=blue,citecolor=blue,pdfpagemode=UseNone,pdfstartview=FitBH]{hyperref}

\newcommand{\tcb}[1]{\textcolor{blue}{#1}}

\begin{document}

\makeatletter\renewcommand{\ps@plain}{%
\def\@evenhead{\hfill\itshape\rightmark}%
\def\@oddhead{\itshape\leftmark\hfill}%
\renewcommand{\@evenfoot}{\hfill\small{--~\thepage~--}\hfill}%
\renewcommand{\@oddfoot}{\hfill\small{--~\thepage~--}\hfill}%
}\makeatother\pagestyle{plain}

\preprint{\textit{Preprint: \today, \now.}} 

\title{Nodeless superconductivity and preserved time-reversal symmetry\\
in the noncentrosymmetric \texorpdfstring{Mo$_3$P}{Mo3P} superconductor}
\author{T.\ Shang}\email[Corresponding authors:\\]{tian.shang@psi.ch}
\affiliation{Laboratory for Multiscale Materials Experiments, Paul Scherrer Institut, Villigen CH-5232, Switzerland}
\author{J.\ Philippe}
\affiliation{Laboratorium f\"ur Festk\"orperphysik, ETH Z\"urich, CH-8093 Zurich, Switzerland}
\author{J.\ A.\ T.\ Verezhak}
\affiliation{Laboratory for Muon-Spin Spectroscopy, Paul Scherrer Institut, CH-5232 Villigen PSI, Switzerland}
\author{Z. Guguchia}
\affiliation{Laboratory for Muon-Spin Spectroscopy, Paul Scherrer Institut, CH-5232 Villigen PSI, Switzerland}
\author{J.\ Z.\ Zhao}
\affiliation{Co-Innovation Center for New Energetic Materials, Southwest University of Science and Technology, Mianyang, 621010, People's Republic of China} 
\author{L.-J.\ Chang}
\affiliation{Department of Physics, National Cheng Kung University, Tainan 70101, Taiwan}
\author{M.~K.~Lee}
\affiliation{Department of Physics, National Cheng Kung University, Tainan 70101, Taiwan}
\author{D.~J.~Gawryluk}\thanks{On leave from Institute of Physics, Polish Academy of Sciences, Aleja Lotnikow 32/46, PL-02-668 Warsaw, Poland.}
\affiliation{Laboratory for Multiscale Materials Experiments, Paul Scherrer Institut, Villigen CH-5232, Switzerland}
\author{E.\ Pomjakushina}
\affiliation{Laboratory for Multiscale Materials Experiments, Paul Scherrer Institut, Villigen CH-5232, Switzerland}
\author{M.\ Shi}
\affiliation{Swiss Light Source, Paul Scherrer Institut, Villigen CH-5232, Switzerland}
\author{M.\ Medarde}
\affiliation{Laboratory for Multiscale Materials Experiments, Paul Scherrer Institut, Villigen CH-5232, Switzerland}
%
%
\author{H.-R.\ Ott}
\affiliation{Laboratorium f\"ur Festk\"orperphysik, ETH Z\"urich, CH-8093 Zurich, Switzerland}
\affiliation{Paul Scherrer Institut, CH-5232 Villigen PSI, Switzerland}
\author{T.\ Shiroka}
\affiliation{Laboratorium f\"ur Festk\"orperphysik, ETH Z\"urich, CH-8093 Zurich, Switzerland}
\affiliation{Paul Scherrer Institut, CH-5232 Villigen PSI, Switzerland}

\begin{abstract}
We report a comprehensive study of the noncentrosymmetric 
superconductor Mo$_3$P. Its bulk superconductivity, with $T_c = 5.5$\,K, 
was characterized via electrical resistivity, magnetization, and heat-capacity 
measurements, while its microscopic electronic properties were investigated 
by means of muon-spin ro\-ta\-tion/re\-la\-xa\-tion ($\mu$SR) and nuclear 
magnetic resonance (NMR) techniques. In the normal state, NMR relaxation 
data indicate an almost ideal metallic behavior, confirmed by 
band-structure calculations, which suggest a relatively high electron 
density of states, dominated by the Mo $4d$-orbitals.
The low-temperature superfluid density, determined via transverse-field 
$\mu$SR and electronic specific heat, suggest a fully-gapped 
superconducting state in Mo$_3$P, with $\Delta_0= 0.83$\,meV, the same 
as the BCS gap value in the weak-coupling case, and a zero-temperature 
magnetic penetration depth $\lambda_0 = 126$\,nm. The absence of spontaneous 
magnetic fields below the onset of superconductivity, as determined from 
zero-field $\mu$SR measurements, indicates a preserved time-reversal 
symmetry in the superconducting state of Mo$_3$P and, hence, 
spin-singlet pairing. 
\end{abstract}


\maketitle\enlargethispage{3pt}

\vspace{-5pt}
\section{\label{sec:Introduction}Introduction}\enlargethispage{8pt}
Non\-cen\-tro\-sym\-met\-ric superconductors (NCSCs) belong to a class 
of materials that miss a key symmetry, such as parity~\cite{Bauer2012}. 
In NCSCs the lack of inversion symmetry of the crystal lattice often 
induces an antisymmetric spin-orbit coupling (ASOC), which lifts the 
degeneracy of the conduction-band electrons and splits the Fermi surface. 
Consequently, both intra- and inter-band Cooper pairs can be formed. 
The admixture of spin-singlet and spin-triplet pairing in NCSCs is 
determined by the strength of the ASOC and by other microscopic 
parameters~\cite{Bauer2012,Smidman2017}. Cooper pairs with a spin-triplet 
pairing have a non\-ze\-ro spin, which implies a weak moment in the 
superconducting state and a breaking of the time-reversal symmetry (TRS). 
This is the case of, e.g., Sr$_2$RuO$_4$ and UPt$_3$, known to exhibit 
triplet superconductivity~\cite{Ishida1998, Tou1998, Mackenzie2003,Joynt2002}, 
and where TRS breaking in the SC state has been confirmed by measurements 
of both zero-field muon-spin relaxation/rotation ($\mu$SR) and 
polar Kerr effect~\cite{Luke1993,Luke1998,Xia2006,Schemm2014}.

Some NCSCs are known to show a broken TRS upon the onset of superconductivity. 
Examples include LaNiC$_2$~\cite{Hillier2009}, La$_7$Ir$_3$~\cite{Barker2015}, 
and some Re-based binary alloys Re$T$ ($T$ -- transition metal, e.g., 
Ti, Zr, Nb, Hf)~\cite{Singh2014,Singh2017,Shang2018,TianReNb2018}.
In the weak SOC limit, TRS breaking can be achieved via nonunitary triplet 
pairing, such as in non-centrosymmetric LaNiC$_2$~\cite{Hillier2009,Quintanilla2012} 
and centrosymmetric LaNiGa$_2$~\cite{Hillier2012,Weng2016}. In case of 
Re$T$ alloys, where a strong SOC is present, the $T_d$ point group has 
several irreducible representations with dimension larger than 1. 
Therefore, they can support TRS breaking with singlet-, triplet-, or 
admixed pairing. For Re$T$, there are a number of possible TRS-breaking 
states~\cite{TianReNb2018}, however, all such states have symmetry-constrained point- or line nodes 
in the gap, inconsistent with the experimentally observed 
nodeless-gap~\cite{Singh2014,Singh2017,Shang2018,TianReNb2018}. 
To explain a fully-gapped superconducting state exihibiting broken TRS, 
a model employing loop-Josephson currents (LJC) was recently proposed~\cite{Ghosh2018}. 
This model is based on an on-site-, intra-orbital-, singlet-pairing SC 
state, where a phase shift among the on-site singlet pairs gives rise 
to the LJC within a unit cell. Such currents can produce weak internal 
magnetic fields, which break the TRS and can, in principle, be detected 
experimentally.

Despite numerous examples of NCSCs, to date only a few of them are known 
to break TRS in their superconducting state. The causes of such a 
selective TRS breaking remain largely unknown. Consequently, comprehensive 
studies of other NCSCs, such as Mo$_3$P reported here, may help to 
identify the origin of this behavior. 
Superconductivity in Mo$_3$P was first reported in 1954~\cite{Matthias1954} 
and confirmed ten years later \cite{Blaugher1965}. In that same year, 
also the Mo$_3$P crystal structure was determined \cite{Sellberg1965}. 
To our knowledge, none of these early works had a follow-up regarding 
the characterization of the superconducting properties of Mo$_3$P.

In this paper, we report on an extensive study of the Mo$_3$P physical 
properties, in the normal and superconducting state, by means of 
electrical resistivity, magnetization, thermodynamic, muon-spin relaxation 
($\mu$SR) and nuclear magnetic resonance (NMR) methods. In addition, we 
also present theoretical density functional theory (DFT) band-structure 
calculations. Despite a non\-cen\-tro\-sym\-met\-ric crystal 
structure, Mo$_3$P is shown to be a moderately 
correlated electron material, which adopts a fully-gapped, spin-singlet 
superconducting state with preserved TRS.

\section{Experimental details\label{sec:details}}\enlargethispage{8pt}
Polycrystalline samples of Mo$_3$P were prepared via solid-state 
reaction, where high-purity (99.999\%) Mo powders and P pieces were 
mixed in a stoichiometric ratio and made to react at 1100\,$^\circ$C 
for 48 hours. The resulting powders were thoroughly grounded and 
pressed into pellets and then sintered at 1100\,$^\circ$C for over 
48 hours. The final product had a silvery color, indicative of good 
metallicity. Room-temperature x-ray powder diffraction (XRD) measurements 
were performed by using a Bruker D8 diffractometer with Cu K$\alpha$ 
radiation. The magnetic susceptibility, electrical resistivity, and 
specific-heat measurements were performed on a 7-T Quantum Design Magnetic 
Property Measurement System (MPMS-7) and a 14-T Physical Property 
Measurement System (PPMS-14) equipped with a dilution refrigerator (DR). 
The $\mu$SR measurements were carried out at the GPS and HAL-9500 spectrometers 
of the Swiss muon source at Paul Scherrer Institut, Villigen, Switzerland~\cite{Amato2017}. 
For the low-temperature HAL measurements, the samples were mounted on a 
silver plate using diluted GE varnish. The $\mu$SR data were analysed by 
means of the \texttt{musrfit} software package \cite{Suter2012}. 

The $^{31}$P NMR measurements, including lineshapes and spin-lattice 
relaxation times were performed on Mo$_3$P powder in two external magnetic 
fields, 0.503 and 7.067\,T. This allowed us to compare the NMR parameters 
in the superconducting and the normal state, respectively. The NMR signals 
were monitored by means of standard spin-echo sequences, consisting in 
$\pi/2$ and $\pi$ pulses of 2.7 and 5.4\,$\mu$s, with recycling delays 
ranging from 0.25 to 20\,s, in a temperature range 1.95 to 295\,K. The 
lineshapes were obtained via fast Fourier transform (FFT) of 
the echo signal. Spin-lattice relaxation times $T_1$ were measured 
via the inversion recovery method, using a $\pi$--$\pi/2$--$\pi$ pulse sequence.

The electronic structure of Mo$_3$P was calculated by using the 
full-potential linearized augmented plane-wave (LAPW) method, as implemented 
in the \texttt{WIEN2k} package~\cite{Blaha2001}. In particular, we employed 
the generalized gradient approximation (GGA)~\cite{Perdew1996}. The 
Brillouin zone integration was performed on a regular mesh of 
$10 \times 10 \times 10$ points. Muffin-tin radii ($r_\mathrm{mt}$) of 
2.41 and 2.06\,$r_\mathrm{Bohr}$ were chosen for Mo and P atoms, 
respectively. The largest plane-wave vector
was defined by $r_\mathrm{mt}k_\mathrm{max}=8$. The SOC
 was treated by using a second-order variational procedure~\cite{Blaha2001}.

\section{\label{sec:results}Results and discussion}\enlargethispage{8pt} 
\subsection{\label{ssec:structure}Crystal structure}

\begin{figure}[!bht]
	\centering 
	\includegraphics[width=0.48\textwidth,angle=0]{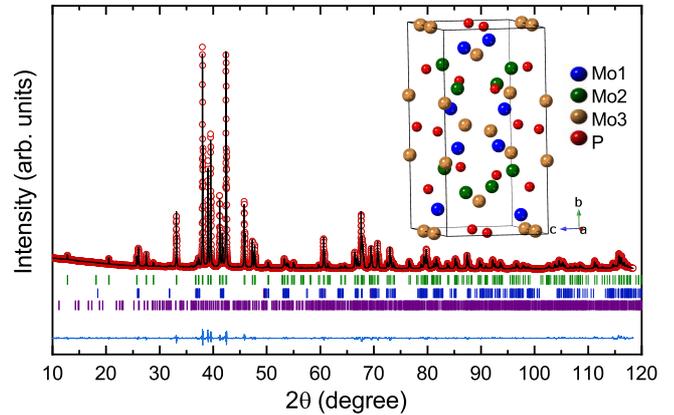} 
	\caption{\label{fig:XRD}Room-temperature x-ray powder diffraction 
	pattern and Rietveld refinement for Mo$_3$P. The open red circles 
	and the solid black line represent the experimental pattern and the 
	Rietveld-refinement profile, respectively. The blue lines at the bottom
	show the residuals, i.e., the difference between calculated and 
	experimental data. The vertical bars mark the calculated Bragg-peak 
positions for Mo$_3$P (green), MoO$_2$ (blue), and Mo$_3$P$_4$ (purple). 
The crystal structure  (unit cell) is shown in the inset.}
\end{figure}

The crystal structure and the purity of Mo$_3$P polycrystalline samples 
were checked via XRD at room temperature. Figure~\ref{fig:XRD} shows a 
representative XRD pattern, analyzed by means of the FullProf Rietveld-analysis 
suite \cite{Carvajal1993}. As previously reported~\cite{Sellberg1965}, 
we confirm that Mo$_3$P crystallizes in the tetragonal noncentrosymmetric 
$\alpha$-V$_3$S-type structure with space group $I\overline{4}2m$ 
(No.\ 121). The refined lattice parameters, $a = b = 9.79094(2)$\,\AA\ 
and $ c = 4.826099(13)$\,\AA, are consistent with the reported values, 
but here established with a substantially improved accuracy. 
According to the refinements in Fig.~\ref{fig:XRD}, tiny amounts of 
Mo$_4$P$_3$ (3.46\%) and MoO$_2$ (4.28\%) phases were also identified. 
The refined Mo$_3$P crystal structure, shown in the inset, comprises 
three different Mo sites and a single P site in the unit cell. 
Table~\ref{tab:atomic} summarizes the atomic positions and the lattice 
parameters. 
\begin{table}[!bht]
	\centering
	\caption{Refined Mo$_3$P crystal-lattice parameters and atomic coordinates, 
	as determined at room temperature. $Z$ represents the number of formula units in the unit cell.
	\label{tab:atomic}} 
	\begin{ruledtabular}
		\begin{tabular}{llcccc}
			Structure & tetragonal, $\alpha$-V$_3$S-type\\
			Space group & $I\overline{4}2m$ (No.\,121)\\ 
		    $Z$         & 8 \\
			$a$(\AA{}) & 9.79094(2)  \\
			$c$(\AA{}) & 4.826099(13)  \\
			$V_\mathrm{cell}$(\AA{}$^3$) &  462.6419(19)  \\ \hline 
	        $R_\mathrm{p} = 2.65$\%,     & 	$R_\mathrm{wp}= 3.70$\%, \quad  $R_\mathrm{exp} = 1.14$\%, \quad	$\chi^2_{r} = 10.5$ \\
		\end{tabular}
		\\ \vspace{8pt}
		Atomic coordinates\\
		\begin{tabular}{lccccc}
			\textrm{Atom}&
			\textrm{Wyckoff}&
			\textrm{Occ.}&
			\textrm{$x$}&
			\textrm{$y$}&
			\textrm{$z$}\\
			\colrule\\
			P      & $8f$ &0.5& 0.2924(3)   & 0           & 0           \\
			Mo$_1$ & $8g$ &0.5& 0.35541(9)  & 0           & 0.5         \\
			Mo$_2$ & $8i$ &0.5& 0.09289(7)  & 0.09289(7)  & 0.2657(3)   \\
			Mo$_3$ & $8i$ &0.5& 0.29871(7)  & 0.29871(7)  & 0.2649(3)   \\
		\end{tabular}
	\end{ruledtabular}
\end{table}
%
\begin{figure}[th]
	\centering
	\includegraphics[width=0.48\textwidth,angle=0]{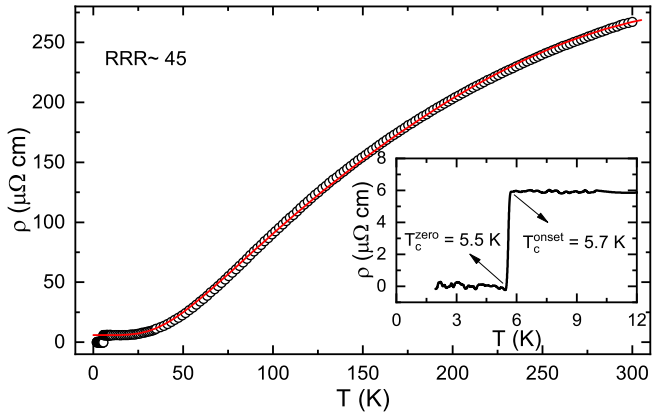}
	\vspace{-2ex}%
	\caption{\label{fig:Rho}Temperature dependence of the Mo$_3$P electrical 
	resistivity. The solid red-line through the data is a fit to Eq.~\eqref{eq:rho_phonon}. 
	The inset shows the enlarged low-temperature data region, highlighting the 
	superconducting transition.}
\end{figure}

\subsection{\label{ssec:rho}Electrical resistivity}
The temperature-dependent electrical resistivity $\rho(T)$ of Mo$_3$P 
was measured from room temperature down to 2\,K. As shown in Fig.~\ref{fig:Rho}, 
the resistivity data reveal a metallic behavior down to the superconducting 
transition at $T_c$. No anomalies associated with structural, magnetic, 
or charge-density-wave transitions could be detected. 
The electrical resistivity in the low-temperature region is shown in the 
inset, where the superconducting transition (with $T_c^\mathrm{onset} = 5.7$\,K 
and $T_c^\mathrm{zero} = 5.5$\,K) is clearly seen. The $\rho(T)$ curve 
below 300\,K can be described by the Bloch-Gr\"{u}neisen-Mott (BGM) 
formula~\cite{Bloch1930,Blatt1968}: 
\begin{equation}
\label{eq:rho_phonon}
\rho(T) = \rho_0 + 4A \left(\frac{T}{\Theta_\mathrm{D}^\mathrm{R}}\right)^5\int_0^{\frac{\Theta_\mathrm{D}^\mathrm{R}}{T}}\!\!\frac{z^2\mathrm{d}z}{(e^z-1)(1-e^{-z})} - \alpha T^3.
\end{equation}
Here the first term $\rho_0$ is the residual resistivity due to the 
scattering of conduction electrons on the static defects of the crystal 
lattice, while the second term describes the electron-phonon scattering,  
with $\Theta_\mathrm{D}^\mathrm{R}$ being the characteristic (Debye) 
temperature and $A$ a coupling constant. The third term represents a 
contribution due to $s$-$d$ interband scattering, $\alpha$ being the 
Mott coefficient~\cite{Mott1958, Mott1964}. 
The fit in Fig.~\ref{fig:Rho} (red-line) results in 
$\rho_0 = 5.92(5)$\,$\mu$$\Omega$cm, $A = 300(7)$\,$\mu$$\Omega$cm, 
$\Theta_\mathrm{D}^\mathrm{R} = 251(3)$\,K, and 
$\alpha = 3.4(1)$\,$\times$10$^{-6}$\,$\mu$$\Omega$cmK$^{-3}$. 
The fairly large residual resistivity ratio (RRR), i.e., 
$\rho(300\,\mathrm{K})/\rho_0 \sim 45$, and the sharp superconducting 
transition temperature ($\Delta T = 0.2$\,K) both indicate a good 
sample quality.

\subsection{\label{ssec:sus} Magnetic susceptibility}
\begin{figure}[th]
	\centering
	\includegraphics[width=0.48\textwidth,angle=0]{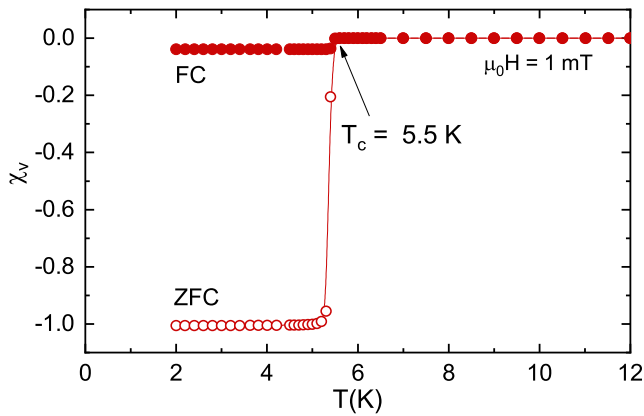}
	\vspace{-2ex}%
	\caption{\label{fig:Chi}Temperature dependence of the Mo$_3$P magnetic 
	susceptibility $\chi_\mathrm{v}$ ($\equiv \chi$), measured using 
	ZFC- and FC protocols. Data were collected in a 1-mT applied field 
	in a 1.8--12\,K temperature range.}
\end{figure}
%

The superconductivity of Mo$_3$P was also probed by magnetic 
susceptibility measurements. The temperature dependence of the magnetic 
susceptibility $\chi(T)$ was determined using both field-cooled (FC) 
and zero-field-cooled (ZFC) protocols in an applied field of 1\,mT. 
As shown in Fig.~\ref{fig:Chi}, $\chi(T)$ data show a superconducting 
transition at $T_c = 5.5$\,K, consistent with the values determined from 
electrical resistivity (Fig.~\ref{fig:Rho}) and heat capacity (see below). 
The splitting of the FC- and ZFC-susceptibilities is a typical feature 
of type-II superconductors with moderate to strong pinning, where the 
magnetic-field flux is pinned upon cooling the material in an applied 
field. Below $T_c$, a ZFC-susceptibility 
value of $\chi_\mathrm{v} \sim -1$ (after accounting for demagnetization factor, 
powder compactness, etc.) indicates bulk superconductivity.

\subsection{\label{ssec:critical_field}Lower and upper critical fields}  
%
\begin{figure}[htp]
	\centering
	\includegraphics[width=0.48\textwidth,angle=0]{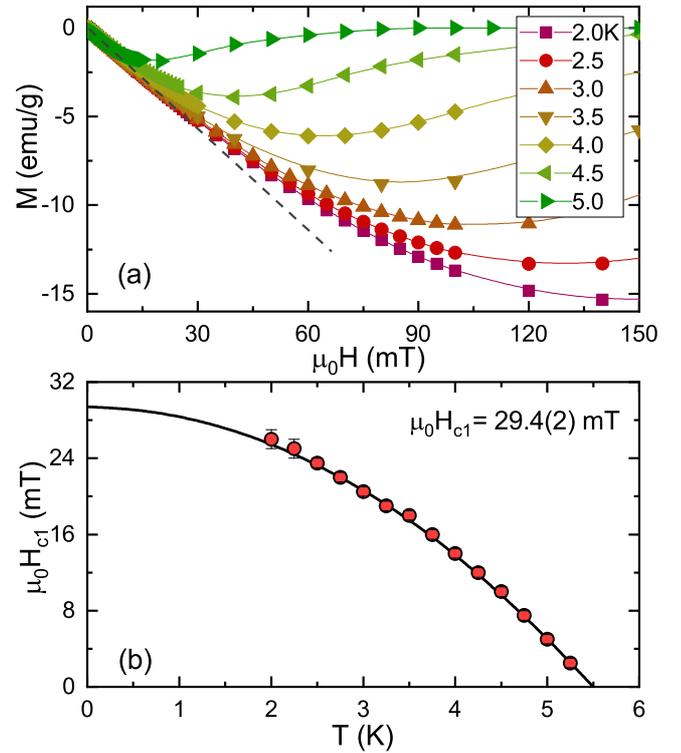}
	\caption{\label{fig:Hc1}(a) Field-dependent magnetization $M(H)$ 
	recorded at various temperatures up to $T_c$. For each temperature, 
	the lower critical field $\mu_{0}H_{c1}$ was determined as the value 
	where $M(H)$ deviates from linearity (dashed line). (b) Estimated 
	lower critical field $\mu_{0}H_{c1}$ as a function of temperature, 
	with the solid-line representing a fit to 
	$\mu_{0}H_{c1}(T) =\mu_{0}H_{c1}(0)[1-(T/T_{c})^2]$.}
\end{figure}
%
To determine the lower critical field $\mu_{0}H_{c1}$ of Mo$_3$P, its 
field-dependent magnetization $M(H)$ was measured at various temperatures 
up to $T_c$, as shown in Fig.~\ref{fig:Hc1}(a). The $M(H)$ curves, 
recorded using a ZFC-protocol, show the typical response of a type-II 
superconductor. The estimated $\mu_{0}H_{c1}$ values at different 
temperatures, determined from the deviation of $M(H)$ from linearity, 
are plotted in Fig.~\ref{fig:Hc1}(b). The solid line is a fit to 
$\mu_{0}H_{c1}(T) = \mu_{0}H_{c1}(0)[1-(T/T_{c})^2]$. This provides  
a lower critical field $\mu_{0} H_{c1}(0) = 29.4(2)$\,mT, consistent with 
24.9(5)\,mT, the value calculated from the magnetic penetration depth 
$\lambda(0)$ (see below).

\begin{figure}[htp]
	\centering
	\includegraphics[width=0.5\textwidth,angle= 0]{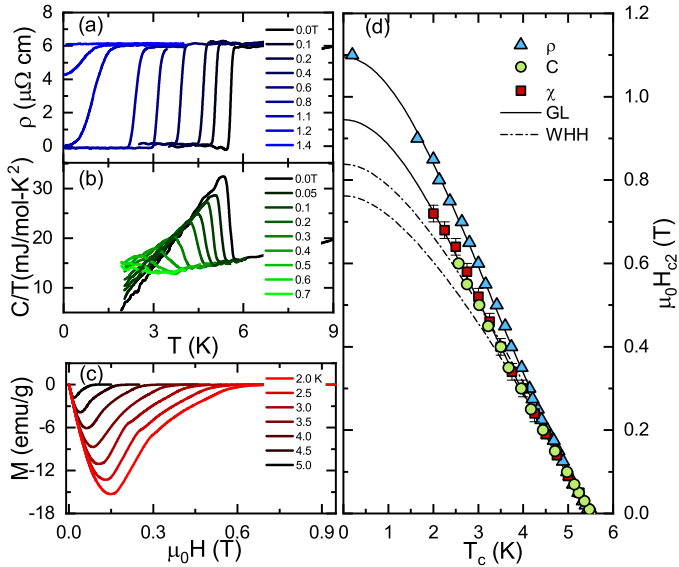}
	\caption{\label{fig:Hc2_determ}Temperature-dependent electrical 
	resistivity $\rho(T,H)$ (a) and specific heat $C/T(T,H)$ (b) for 
	different applied magnetic fields, up to 1.4\,T. In the first case, 
	$T_c$ was defined as the onset of zero resistance. (c) Field-dependent 
	magnetization $M(H,T)$ measured at various temperatures and in 
	applied magnetic fields up to 1.0\,T. From $\rho(T,H)$, $C(T,H)/T$, 
	and $M(H,T)$ one can derive the upper critical field $\mu_{0}H_{c2}$, 
	as shown in (d). The solid lines represent fits to an effective 
	Ginzburg-Landau (GL) model, whereas the dash-dotted lines are fits 
	based on a WHH model without spin-orbit scattering.}
\end{figure}
%

To investigate the behavior of the upper critical field $\mu_0$$H_{c2}$, 
we measured the temperature-dependent electrical resistivity $\rho(T)$ 
and specific heat $C(T)$/$T$ at various applied magnetic fields, as well 
as the field-dependent magnetization $M(H)$ at various temperatures 
up to $T_c$. As shown in Figs.~\ref{fig:Hc2_determ}(a) and (b), the 
superconducting transition, determined from 
$\rho(T)$ and specific heat $C(T)$/$T$ data, 
shifts towards lower temperature upon increasing the magnetic field. 
In zero magnetic field, $T_c = 5.5$\,K determined from $C(T)$/$T$ 
is consistent with the $T_c$ values determined from $\rho(T)$ 
(Fig.~\ref{fig:Rho}) and $\chi(T)$ (Fig.~\ref{fig:Chi}) data. In the 
latter case, as shown in Fig.~\ref{fig:Hc2_determ}(c), the diamagnetic 
signal disappears once the applied magnetic field exceeds the upper 
critical field. 
Figure~\ref{fig:Hc2_determ}(d) summarizes the upper critical fields 
vs.\ the superconducting transition temperatures $T_c$, as derived 
from $\rho(T, H)$, $C(T,H)/T$, and $M(H, T)$ data, respectively. 
The temperature dependence of $\mu_0 H_{c2}(T)$ was analyzed by means 
of the semi-empirical model $\mu_0 H_{c2}(T) = \mu_0 H_{c2}(0)(1-t^2)/(1+t^2)$, 
where $t = T/T_{c}$ is the normalized temperature (see, e.g., \cite{Zhu2008}). 
This model follows from the Ginzburg-Landau (GL) relation between 
$H_{c2}$ and the coherence length $\xi$ (see below), by assuming 
$\xi(t) \propto \sqrt{(1+t^2)/(1-t^2)}$. Although the GL theory is 
strictly valid near $T_c$, the above form of $H_{c2}(T)$ 
has been shown to be satisfied in a wider temperature range.
The solid lines in Fig.~\ref{fig:Hc2_determ}(d) are fits to the GL model, 
which gives $\mu_0 H_{c2}^\mathrm{GL}(0) = 0.94(1)$\,T and 1.09\,T for 
specific heat and magnetization, and for electrical resistivity data, 
respectively. For a comparison, we estimated the upper critical field 
also by means of the Werthamer-Helfand-Hohenberg (WHH) model~\cite{Werthamer1966}. 
The dashed-lines in Fig.~\ref{fig:Hc2_determ}(d) are fits to a WHH model 
without spin-orbital scattering and give $\mu_0 H_{c2}^\mathrm{WHH}(0) = 0.76(1)$\,T 
and 0.83(1)\,T for specific heat and magnetization, and for electrical 
resistivity data, respectively. 

At low fields, both GL and WHH models describe the experimental data very well.\
 At higher applied fields, however, the WHH-type fits deviate 
significantly from the data, clearly providing underestimated 
critical-field values. The remarkable agreement of the less elaborate GL 
model with experimental data is clearly seen in Fig.~\ref{fig:Hc2_determ}(d). 
We note that, while the specific-heat-, magnetization-, and 
electrical-resistivity datasets agree well at low fields ($<$\,0.25\,T), 
at higher fields, the transition temperatures determined from $\rho(T,H)$ 
are systematically higher than those derived from $C(T,H)/T$ and $M(H,T)$ 
data. A similar behavior has also been found in other NCSCs, e.g., 
LaPtSi, BiPd, or La$T$Si$_3$ ($T$ = Pd, Pt, and 
Ir)~\cite{Kneidinger2013,Peets2016,Kimura2016,Anand2014,Smidman2014}, 
as well as in standard superconductors.
Generally, the surface/filamentary superconductivity above bulk $T_c$, or a strong anisotropy of the upper critical field, are proposed as the cause of the differing $T_c$ values.
In order to clarify this issue, studies of single-crystal specimen are highly desirable.

In the GL  
theory of superconductivity, the coherence length 
$\xi$ can be calculated from $\xi$ =  $\sqrt{\Phi_0/2\pi\,H_{c2}}$, where 
$\Phi_0 = 2.07 \times 10^{-3}$\,T~$\mu$m$^{2}$ is the quantum of magnetic 
flux. With a bulk $\mu_{0}H_{c2}(0) = 0.94(1)$\,T, the calculated $\xi(0)$ 
is 18.7(1)\,nm. The magnetic penetration depth $\lambda$ is related to 
the coherence length $\xi$ and the lower critical field $\mu_{0}H_{c1}$ 
via $\mu_{0}H_{c1} = (\Phi_0 /4 \pi \lambda^2)[$ln$(\kappa)+ 0.5]$, where 
$\kappa$ = $\lambda$/$\xi$ is the GL parameter~\cite{Brandt2003}.
By using $\mu_{0}H_{c1} = 29.4$\,mT and $\xi = 18.7$\,nm, the resulting 
magnetic penetration depth $\lambda_\mathrm{GL}$ = 113(1)\,nm, is 
consistent with 126(1) (zero-field extrapolation) and 121(2)\,nm (50\,mT), 
the experimental values from TF-$\mu$SR data (see Sec.~\ref{ssec:TF_muSR} and Table~\ref{tab:parameter}). 
A GL parameter $\kappa \sim 7$, almost ten times larger than the $1/\sqrt{2}$ 
threshold value, clearly confirms that Mo$_3$P is a type-II superconductor.

\subsection{\label{ssec:Cp_zero} Zero-field specific heat}
\begin{figure}[th]
	\centering
	\includegraphics[width=0.45\textwidth,angle=0]{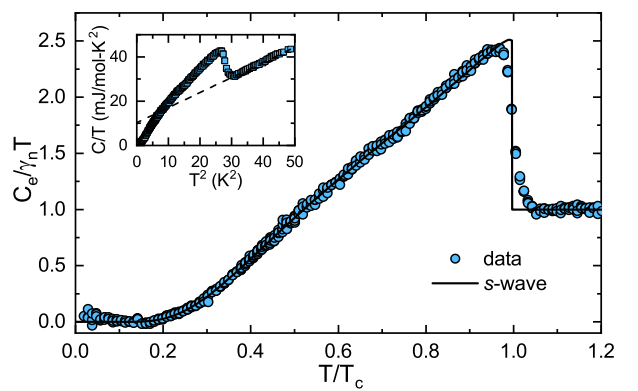}
	\vspace{-2ex}%
	\caption{\label{fig:Cp}Normalized electronic specific heat 
	$C_\mathrm{e}/\gamma_n T$ for Mo$_3$P versus $T/T_c$. Inset: the 
	measured specific heat $C/T$ as a function of $T^2$. The dashed-line 
	in the inset is a fit to $C/T = \gamma + \beta T^2$, while the solid 
	line in the main panel is the electronic specific heat calculated 
	by considering a fully-gapped $s$-wave model.} 
\end{figure}

Specific-heat data 
offer valuable insight into the superconducting 
properties, including the gap value and its symmetry. Therefore, the 
Mo$_3$P specific heat was measured down to 0.1\,K in zero field. As shown 
in Fig.~\ref{fig:Cp}, there is a clear specific-heat jump at $T_c$, 
indicating a bulk superconducting transition. The electronic specific 
heat $C_\mathrm{e}$/$T$ was obtained by subtracting the phonon contribution 
from the experimental data. As shown in the inset of Fig.~\ref{fig:Cp}, 
the normal-state specific heat of Mo$_3$P is fitted to 
$C/T = \gamma_\mathrm{n} + \beta T^2$, where $\gamma_\mathrm{n}$ is 
the electronic specific-heat coefficient and $\beta T^2$ is the phonon 
contribution to the specific heat. The derived values are 
$\gamma_\mathrm{n} = 10.3(4)$\,mJ/mol-K$^2$ and $\beta = 0.67(3)$\,mJ/mol-K$^4$. 
The Debye temperature $\Theta_\mathrm{D}^\mathrm{C}$ can be calculated 
by using $\Theta_\mathrm{D}^\mathrm{C} = (12\pi^4\,Rn/5\beta)^{1/3}$, 
where $R = 8.314$\,J/mol-K is the molar gas constant and $n = 4$ is the 
number of atoms per formula unit. The resulting $\Theta_\mathrm{D}^\mathrm{C} = 225(3)$\,K 
(extracted from low-$T$ data) is consistent with the value derived from 
electrical-resistivity data (see Fig.~\ref{fig:Rho}). The density of 
states (DOS) at the Fermi level $N(\epsilon_\mathrm{F})$ was evaluated 
from the expression 
$N(\epsilon_\mathrm{F}) = 3\gamma_\mathrm{n}/(2\pi^2 k_\mathrm{B}^2) = 2.2(1)$\,states/eV-f.u.\ 
(accounting for spin de\-ge\-ne\-ra\-cy)~\cite{Kittel2005}, where 
$k_\mathrm{B}$ is the Boltzmann constant. The electron-phonon coupling 
constant $\lambda_\mathrm{ep}$, a measure of the attractive interaction 
between electrons due to phonons, was estimated from the 
$\Theta_\mathrm{D}^\mathrm{C}$ and $T_c$ values by applying the semi-empirical 
McMillan formula~\cite{McMillan1968}:
\begin{equation}
\label{eq:lambda_ep}
\lambda_\mathrm{ep}=\frac{1.04+\mu^{\star}\,\mathrm{ln}(\Theta_\mathrm{D}/1.45\,T_c)}{(1-0.62\,\mu^{\star})\mathrm{ln}(\Theta_\mathrm{D}/1.45\,T_c)-1.04}.
\end{equation}
%
%
The Coulomb pseudopotential 
$\mu^{\star}$, usually lying in the 0.09--0.18 
range, was here fixed to 0.13, a commonly used value for 
transition metals. From Eq.~\eqref{eq:lambda_ep} we obtain 
$\lambda_\mathrm{ep} = 0.72(1)$ for Mo$_3$P, almost twice the 
reported value for the elemental Mo (0.41) 
\cite*{Boughton1970,Shang2019}. Finally, the band-structure density of 
states $N_\mathrm{band}(\epsilon_\mathrm{F})$ can be estimated from the relation 
$N_\mathrm{band}(\epsilon_\mathrm{F}) = N(\epsilon_\mathrm{F})/(1 + \lambda_\mathrm{ep}$)~\cite{Kittel2005}, 
which gives $N_\mathrm{band}(\epsilon_\mathrm{F})$ = $1.28(1)$\,states/eV-f.u.

The electronic specific heat divided by the electronic specific-heat 
coefficient, i.e., $C_\mathrm{e} / \gamma_\mathrm{n} T$, is shown in the 
main panel of Fig.~\ref{fig:Cp} as a function of the reduced temperature. 
The temperature-independent behavior of $C_\mathrm{e} / \gamma_\mathrm{n} T$ 
at low-$T$ suggests a fully-gapped superconducting state in Mo$_3$P. 
The temperature-dependent superconducting-phase contribution to the 
entropy was calculated by means of \cite{Tinkham1996}:
\begin{equation}
S(T) = -\frac{6\gamma_\mathrm{n}}{\pi^2 k_\mathrm{B}} \int^{\infty}_0 [f\mathrm{ln}f+(1-f)\mathrm{ln}(1-f)]\,\mathrm{d}\epsilon,
\end{equation}
where $f = (1+e^{E/k_\mathrm{B}T})^{-1}$ is the Fermi function, $\Delta_0$ 
is the SC gap value at 0\,K, and $E(\epsilon) = \sqrt{\epsilon^2 + \Delta^2(T)}$ 
is the excitation energy of quasiparticles, with $\epsilon$ the electron 
energies measured relative to the chemical potential (Fermi 
energy)~\cite{Padamsee1973,Tinkham1996}. $\Delta(T)$ is the same as in 
Sec.~\ref{ssec:TF_muSR}. The temperature-dependent electronic specific 
heat in the superconducting state can be calculated from 
$C_\mathrm{e} = T \frac{dS}{dT}$. The solid line in Fig.~\ref{fig:Cp} 
represents a fit with a fully-gapped $s$-wave model. The resulting 
superconducting gap $\Delta_0 = 0.82(1)$\,meV is consistent with the 
$\mu$SR results (see Fig.~\ref{fig:lambda}). The Mo$_3$P gap value 
$\Delta_0$, as derived from both specific-heat- and TF-$\mu$SR data, is 
comparable to the expected weak-coupling BCS value 0.80\,meV, thus indicating 
weakly-coupled superconducting pairs in Mo$_3$P. Furthermore, also the 
specific-heat discontinuity at $T_c$, i.e., $\Delta C/\gamma_\mathrm{n} T_{c} = 1.44$, 
is fully consistent with the conventional BCS value of 1.43.

\subsection{\label{ssec:TF_muSR}Transverse-field \texorpdfstring{$\mu$SR}{MuSR}}
\begin{figure}[!htp]
	\centering
	\includegraphics[width=0.50\textwidth,angle= 0]{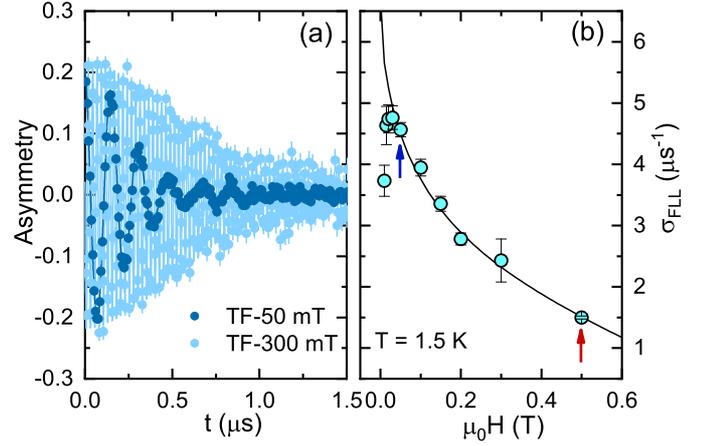}
	\caption{\label{fig:TF-muSR_H}(a) Time-domain TF-$\mu$SR spectra of 
	Mo$_3$P measured in its superconducting state (at $T = 1.5$\,K) at 
	50 and 300\,mT. (b) Field-dependent Gaussian relaxation rate 
	$\sigma_\mathrm{FLL}(H)$. The arrows indicate the field values (50 
	and 500\,mT) used in the temperature-dependent TF-$\mu$SR studies. 
	The solid line is a fit to Eq.~\eqref{eq:TF_muSR_H}.}
\end{figure}

To investigate the superconducting properties of Mo$_3$P at a microscopic 
level, we carried out $\mu$SR measurements in applied transverse fields 
(TF). The optimal field value for such experiments was determined via 
a preliminary field-dependent $\mu$SR depolarization-rate measurement 
at 1.5\,K. To track the additional field-distribution broadening due 
to the flux-line-lattice (FLL) in the mixed superconducting state, a 
magnetic field (up to 500\,mT) was applied in the normal state and then 
the sample was cooled down to 1.5\,K. Figure~\ref{fig:TF-muSR_H}(a) 
show the TF-spectra in an applied field of 50 and 300\,mT, respectively. 
The solid lines represent fits using the same model as described in 
Eq.~\eqref{eq:TF_muSR} below. 
%
%
The resulting effective Gaussian relaxation rate $\sigma_\mathrm{FLL}(H)$ 
is summarized in Fig.~\ref{fig:TF-muSR_H}(b). Above the lower critical 
field $\mu_{0}H_{c1}$ (29.4\,mT), the 
relaxation rate decreases 
continuously. By considering the decrease of the inter vortex distance 
with the field and the vortex-core effects, a field of 50\,mT was chosen 
for the temperature-dependent TF-$\mu$SR studies. For a comparison, 
the TF-$\mu$SR relaxation was also measured in a field of 500\,mT. 
The field dependence of the Gaussian relaxation rate was analyzed 
following Eq.~\eqref{eq:TF_muSR_H} (see details below). The derived 
zero-temperature $\mu_0$$H_{c2}$ = 0.96(1)\,T and magnetic penetration 
depth $\lambda_0 = 126(1)$\,nm are consistent with the temperature-dependent 
50-mT results (see Table~\ref{tab:parameter}). 

\begin{figure}[!thp]
	\centering
	\includegraphics[width=0.49\textwidth,angle= 0]{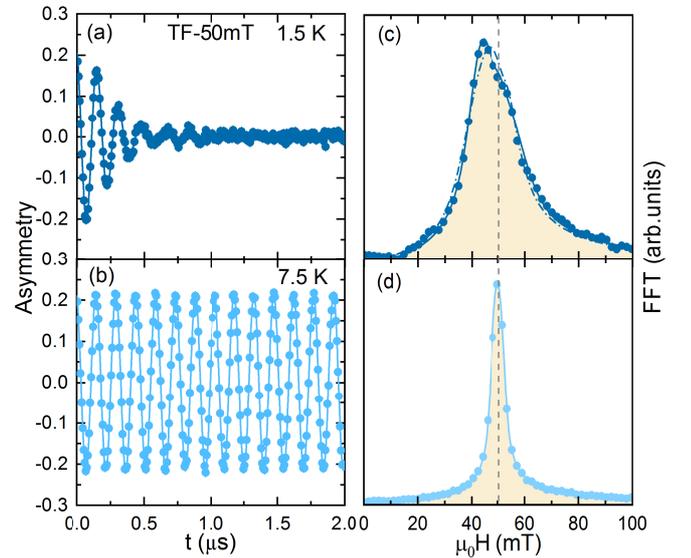}
	\caption{\label{fig:TF-muSR_T}TF-$\mu$SR time-dependent 
		spectra, collected at 1.5\,K (a) and 7.5\,K (b) 
		in an applied field of 50\,mT. Fourier transforms of the above 
		time spectra at 1.5\,K (c) and 7.5\,K (d). The solid lines 
		are fits to Eq.~(\ref{eq:TF_muSR}) using two oscillations, 
		while the dash-dotted line in (c) represents a fit with a 
		single oscillation. The vertical dashed line indicates the 
		applied magnetic field. Note the clear 
		diamagnetic shift below $T_{c}$ in (c).}
\end{figure}

The TF-$\mu$SR time spectra were collected at various temperatures up 
to $T_c$, following a field-cooling protocol on both GPS and HAL 
instruments. Figures~\ref{fig:TF-muSR_T}(a) and (b) show two representative 
TF-$\mu$SR spectra collected above (7.5\,K) and below $T_c$ (1.5\,K) on 
GPS. The enhanced depolarization rate below $T_c$ reflects the 
inhomogeneous field distribution due to the FLL, which causes the 
additional distribution broadening in the mixed state [see 
Fig.~\ref{fig:TF-muSR_T}(c)]. The time evolution of the $\mu$SR-asymmetry 
can be modelled by: 
\begin{equation}
\label{eq:TF_muSR}
A_\mathrm{TF}(t) = \sum\limits_{i=1}^n A_i \cos(\gamma_{\mu} B_i t + \phi) e^{- \sigma_i^2 t^2/2} +
A_\mathrm{bg} \cos(\gamma_{\mu} B_\mathrm{bg} t + \phi).
\end{equation}
Here $A_i$ and $A_\mathrm{bg}$ represent the initial muon-spin asymmetries 
for muons implanted in the sample and sample holder, respectively, with 
the latter not undergoing any depolarization. $B_i$ and $B_\mathrm{bg}$ 
are the local fields sensed by implanted muons in the sample and sample 
holder, $\gamma_{\mu} = 2\pi \times 135.53$\,MHz/T is the muon 
gyromagnetic ratio, $\phi$ is the shared initial phase, and $\sigma_i$ 
is a Gaussian relaxation rate of the $i$th component. The number of 
required components is material dependent, generally in the $1 \leq n \leq 5$ 
range. For superconductors with a large $\kappa$ ($\gg 1$) value, the 
magnetic penetration depth is much larger than the coherence length. 
Hence, the field distribution is narrow and a single-oscillating component 
is sufficient to describe $A(t)$. In case of a small $\kappa$ 
($\gtrsim 1/\sqrt{2}$) value, the magnetic penetration depth is 
comparable to the coherence length. This implies a broad field distribution, 
in turn requiring multiple oscillations to describe $A(t)$. 

\begin{figure}[!thp]
	\centering
	\includegraphics[width=0.49\textwidth,angle= 0]{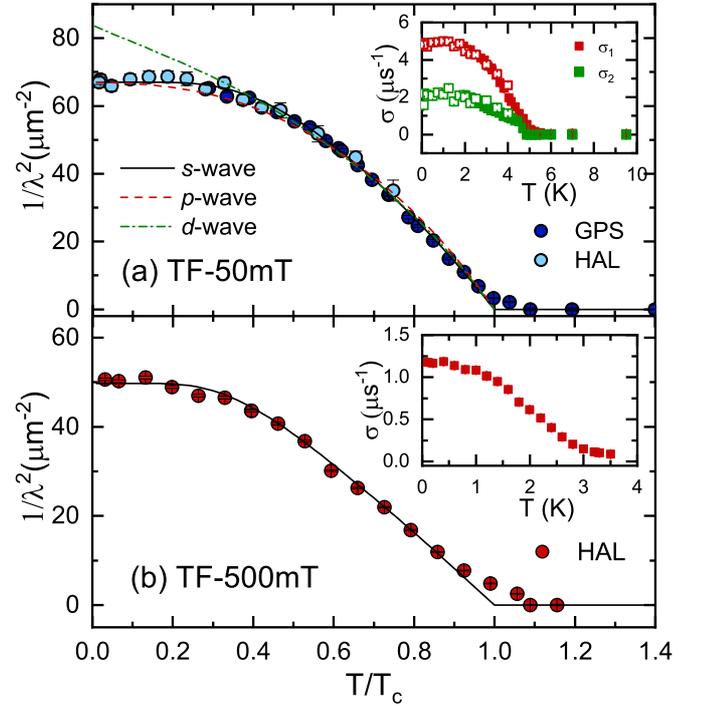}
	\caption{\label{fig:lambda}Superfluid density vs.\ temperature, as 
	determined from TF-$\mu$SR measurements in an applied magnetic field 
	of 50\,mT (a) and 500\,mT (b). The insets show the temperature 
	dependence of the muon-spin relaxation rate $\sigma(T)$. Two components 
	are required to describe the 50-mT experimental data [see details in 
	Fig.~\ref{fig:TF-muSR_T}(c)]. The different lines represent fits to 
	various models, including $s$-, $p$-, and $d$-wave pairing (see text 
	for details).}
\end{figure}

Figures~\ref{fig:TF-muSR_T}(c) and (d) show the FFT spectra of the TF-$\mu$SR time-domain asymmetries at 1.5\,K and 
7.5\,K. Solid lines represent fits to Eq.~\eqref{eq:TF_muSR} using two 
oscillations (i.e., $n = 2$) in the superconducting phase and one 
oscillation in the normal phase. In the 1.5\,K case, a single-component 
oscillation [dash-dotted line in panel (c)], is clearly unable to fit 
the $T < T_{c}$ data. The derived Gaussian relaxation rates as a function 
of temperature are summarized in the insets of Fig.~\ref{fig:lambda}. 
Above $T_c$ the relaxation rate is small and temperature-independent, but 
below $T_c$ it starts to increase due to the onset of the FLL and the 
increase in superfluid density. Unlike the TF-$\mu$SR spectra collected 
at 50\,mT, the 500-mT datasets can be described by a single oscillation. 
This reflects the narrower internal-field distribution for applied fields 
comparable to $H_{c2}$. In case of multi-component oscillations (at 50\,mT), 
the first-term in Eq.~\eqref{eq:TF_muSR} describes the field distribution 
as the sum of $n$ Gaussian relaxations~\cite{Maisuradze2009}:
\begin{equation}
\label{eq:TF_muSR_2}
P(B) = \gamma_{\mu} \sum\limits_{i=1}^2 \frac{A_i}{\sigma_i} \mathrm{exp}\left[-\frac{\gamma_{\mu}^2(B-B_i)^2}{2\sigma_i^2}\right].
\end{equation}
Then, the first and second moments of the field distribution in the 
sample were calculated by: 
\begin{equation}
\label{eq:1st_moment}
	\langle B \rangle =  \sum\limits_{i=1}^2 \frac{A_i B_i}{A_\mathrm{tot}},\quad \mathrm{and}
\end{equation}
\begin{equation}
\label{eq:2nd_moment}
 \langle B^2 \rangle = \frac{\sigma_\mathrm{eff}^2}{\gamma_\mu^2} = \sum\limits_{i=1}^2 \frac{A_i}{A_\mathrm{tot}}\left[\frac{\sigma_i^2}{\gamma_{\mu}^2} - \left(B_i - \langle B \rangle\right)^2\right],
\end{equation}
where $A_\mathrm{tot} = \sum_{i=1}^2 A_i$. In the superconducting state, 
the measured Gaussian relaxation rate includes contributions from both a 
temperature-independent relaxation due to nuclear moments ($\sigma_\mathrm{n}$) 
and the FLL ($\sigma_\mathrm{FLL}$). The FLL-related relaxation can be 
extracted by subtracting the nuclear contribution according to 
$\sigma_\mathrm{FLL}$ = $\sqrt{\sigma_\mathrm{eff}^{2} - \sigma^{2}_\mathrm{n}}$.
Since $\sigma_\mathrm{FLL}$ is directly related to the magnetic penetration 
depth and the superfluid density ($\sigma_\mathrm{FLL} \propto 1/\lambda^2$), 
the superconducting gap value and its symmetry can be determined from 
the measured $\sigma_\mathrm{FLL}$$(T)$. 

%

For small applied magnetic fields ($H_\mathrm{appl}/H_{c2}\ll\,1$) and  
large $\kappa$ values ($\kappa \gg 1$), $\sigma_\mathrm{FLL}$ is 
field-in\-de\-pend\-ent and proportional to $\lambda^{-2}$~\cite{Barford1988,Brandt2003}. 
In the Mo$_3$P case, however, the upper critical field $\mu H_{c2}$ is 
relatively small (0.94\,T) compared to the applied TF-field (50 and 500\,mT). 
Therefore, to extract the penetration depth from the measured $\sigma_\mathrm{FLL}$, 
the effects of overlapping vortex cores with increasing field have to 
be considered, requiring the following expression for calculating the 
magnetic penetration depth $\lambda$~\cite{Barford1988,Brandt2003}:
\begin{equation}
\label{eq:TF_muSR_H}
\sigma_\mathrm{FLL} = 0.172 \frac{\gamma_{\mu} \Phi_0}{2\pi}(1-h)[1+1.21(1-\sqrt{h})^3]\lambda^{-2}, 
\end{equation} 
where $h = H_\mathrm{appl}/H_\mathrm{c2}$, with $H_\mathrm{appl}$ the 
applied magnetic field. The above expression is valid for type-II 
superconductors with $\kappa \ge 5$ in the $0.25/\kappa^{1.3} \lesssim h \le$ 1 
field range. With $\kappa \sim 6$ and $h = 0.053$ (TF-50\,mT) and 0.53 
(TF-500\,mT), Mo$_3$P fulfills the above conditions. 

By using Eq.~\eqref{eq:TF_muSR_H}, we could calculate the inverse-square 
of the magnetic penetration depth, whose temperature dependence is shown 
in Fig.~\ref{fig:lambda}. To gain insight into the superconducting 
pairing symmetry in Mo$_3$P, the temperature-dependent superfluid density 
$\rho_\mathrm{sc}(T)$ [$\lambda^{-2}(T) \propto \rho_s(T)$] was further 
analyzed by means of different models, generally described by:
\begin{equation}
\label{eq:rhos}
\rho_\mathrm{sc}(T) = 1 + 2\, \Bigg{\langle} \int^{\infty}_{\Delta_k} \frac{E}{\sqrt{E^2-\Delta_k^2}} \frac{\partial f}{\partial E} \mathrm{d}E \Bigg{\rangle}_\mathrm{FS},
\end{equation}
where $\Delta_k$ is an angle-dependent gap function, 
$f = (1+e^{E/k_\mathrm{B}T})^{-1}$ is the Fermi function, and 
$\langle \rangle_\mathrm{FS}$ represents an average over the Fermi 
surface \cite{Tinkham1996}. The gap function can be written as 
$\Delta_\mathrm{k}(T) = \Delta(T) g_{k}$, where $\Delta$ is the 
maximum gap value and $g_\mathrm{k}$ is the angular dependence of the gap, 
equal to 1, $\cos2\psi$, and $\sin\theta$ for an $s$-, $d$-, and $p$-wave 
model, respectively. Here $\psi$ and $\theta$ are azimuthal angles.
The temperature dependence of the gap is assumed to follow 
$\Delta(T) = \Delta_0 \mathrm{tanh} \{ 1.82[1.018(T_c/T-1)]^{0.51} \}$ 
\cite{Carrington2003}, where $\Delta_0$, the gap value at zero temperature, 
is the only adjustable parameter. Note that the function $\Delta(T)$ is 
practically mo\-del-in\-de\-pend\-ent. 

Three different models, including $s$-, $d$-, and $p$ waves, were used 
to describe the temperature-dependent superfluid density $\lambda^{-2}$$(T)$ 
measured in an applied field of 50\,mT [Fig.~\ref{fig:lambda}(a)], with 
the first two being singlet-pairing and the last one triplet-paring. 
For the $s$- and $p$-wave model, the estimated gap values are 0.83(1) 
and 1.12(1)\,meV, respectively, with the shared zero-temperature magnetic 
penetration depth $\lambda_\mathrm{0} = 121(2)$\,nm; while for the 
$d$-wave model, the estimated $\lambda_\mathrm{0}$ and gap values are 
109(2)\,nm and 1.10(1)\,meV. As can be seen in Fig.~\ref{fig:lambda}(a), 
the temperature-independent behavior of $\lambda^{-2}(T)$ below $1/3\,T_c$ 
strongly suggests a nodeless superconductivity in Mo$_3$P. Therefore, 
$\lambda^{-2}(T)$ is clearly more consistent with a single fully-gapped 
$s$-wave model. In case of $d$- or $p$-wave models, a less-good agreement 
with the measured $\lambda^{-2}$ values is found, especially 
at low temperature. Such conclusions are further supported by the low-$T$ 
specific-heat data shown in Fig.~\ref{fig:Cp}. Also in the TF-500\,mT 
case [see Fig.~\ref{fig:lambda}(b)], the $\lambda^{-2}(T)$ dependence 
is consistent with an $s$-wave model, leading to a $\lambda_\mathrm{0} = 141(2)$\,nm 
and to a gap value $\Delta_{0} = 1.57\,k_\mathrm{B}T_c = 0.41(1)$\,meV at 0.5\,T.

\subsection{\label{ssec:ZF_muSR}Zero-field \texorpdfstring{$\mu$SR}{MuSR}}
\begin{figure}[ht]
	\centering
	\includegraphics[width=0.49\textwidth,angle=0]{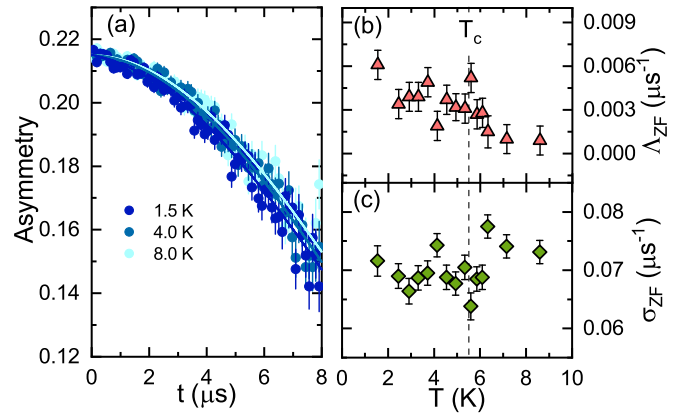}
	\vspace{-2ex}%
	\caption{\label{fig:ZF_muSR}(a) Representative ZF-$\mu$SR spectra for 
	Mo$_3$P in the superconducting (1.5 and 4\,K) and the normal state (8\,K). 
	The solid lines are fits to Eq.~(\ref{eq:KT_and_electr}), as described 
	in the text. Temperature dependence of the Lorentzian 
	relaxation rate $\Lambda_\mathrm{ZF}$ (b) and Gaussian relaxation rate $\sigma_\mathrm{ZF}$ (c). 
	The dashed line, which marks the bulk $T_c$ at 5.5\,K, is a guide to the eyes. None of the 
	reported fit parameters show distinct anomalies across $T_c$.}
\end{figure}

We performed also ZF-$\mu$SR measurements, in order to search for a 
possible TRS breaking in the superconducting state of Mo$_3$P. The 
large muon gyromagnetic ratio, combined with the availability of 100\% 
spin-polarized muon beams, make ZF-$\mu$SR a very sensitive probe 
for detecting small spontaneous magnetic fields. This technique has 
been successfully used to detect the TRS breaking in the superconducting 
states of different types of 
materials~\cite{Hillier2009,Barker2015,Luke1998,aoki2003,Shang2018,TianReNb2018}. 
Normally, in the absence of external fields, the onset of SC does not 
imply changes in the ZF muon-spin relaxation rate. However, if the 
TRS is broken, the onset of tiny spontaneous currents gives rise to 
associated (weak) magnetic fields, readily detected by ZF-$\mu$SR as 
an increase in muon-spin relaxation rate. 
Representative ZF-$\mu$SR spectra collected above (8\,K) and below $T_c$ 
(1.5 and 4\,K) 
are shown in Fig.~\ref{fig:ZF_muSR}. For 
non-magnetic materials, in the absence of applied fields, the relaxation 
is mainly determined by the randomly oriented nuclear moments. 
Consequently, the ZF-$\mu$SR spectra of Mo$_3$P could be modelled by 
means of a combined Lorentzian and Gaussian Kubo-Toyabe relaxation 
function~\cite{Kubo1967, Yaouanc2011}:
\begin{equation}
\label{eq:KT_and_electr}
A_\mathrm{ZF} = A_\mathrm{s}\left[\frac{1}{3} + \frac{2}{3}(1 -
\sigma_\mathrm{ZF}^{2}t^{2} - \Lambda_\mathrm{ZF} t)\,
\mathrm{e}^{\left(-\frac{\sigma_\mathrm{ZF}^{2}t^{2}}{2} - \Lambda_\mathrm{ZF} t\right)} \right] + A_\mathrm{bg}.
\end{equation}
%
%
Here $A_\mathrm{s}$ and $A_\mathrm{bg}$ represent the initial muon-spin 
asymmetries for muons implanted in the sample and sample holder, 
respectively. In polycrystalline samples, the 1/3-nonrelaxing and 
2/3-relaxing components of the asymmetry correspond to the powder 
average of the local internal fields with respect to the initial 
muon-spin orientation. After fixing $A_\mathrm{s}$ to its 
average value of 0.215, the resulting fit parameters are shown in 
Fig.~\ref{fig:ZF_muSR}(b)-(c). The weak Gaussian and Lorentzian 
relaxation rates reflect the absence of electronic magnetic moments and 
the small value of the Mo$_3$P nuclear moments. Although the ZF-$\mu$SR 
spectra in Fig.~\ref{fig:ZF_muSR}(a) show tiny differences between the normal 
and the superconducting state, the resulting $\Lambda_\mathrm{ZF}(T)$ 
and $\sigma_\mathrm{ZF}(T)$ parameters show no distinct changes across 
$T_c$. Note that the tiny increase of $\Lambda_\mathrm{ZF}$ between 
the normal and superconducting state [Fig.~\ref{fig:ZF_muSR}(b)] is 
not related to a TRS-breaking effect, but to the correlated 
decrease of $\sigma_\mathrm{ZF}$ [Fig.~\ref{fig:ZF_muSR}(c)].
The clear lack of an additional relaxation below $T_c$ implies 
that the TRS is preserved in the superconducting state of Mo$_3$P.

\subsection{\label{ssec:NMR}\texorpdfstring{$^{31}$P}{31P} NMR in the normal phase}
NMR is a versatile technique for investigating the electronic properties 
of materials, in particular, their electron correlations, complementary 
to $\mu$SR with respect to probe location. Considering the rather low 
$T_{c}$- and $H_{c2}(0)$ values of Mo$_{3}$P, as well as the detailed 
$\mu$SR study of its superconducting properties (see previous section), 
we mostly employed NMR to investigate the normal-state electronic properties 
of Mo$_{3}$P. To this aim, $^{31}$P NMR measurements were performed at 
7 and 0.5\,T, i.e., above and below $H_{c2}(0)$ (0.94\,T).
For both fields, the $^{31}$P NMR reference frequency \tcb{$\nu_0$} was determined by 
using a solid-state NH$_{4}$H$_{2}$PO$_{4}$ sample, whose Larmor frequency 
coincides with that of phosphoric acid. The $^{31}$P NMR shifts were 
then calculated with respect to this reference.

\begin{figure}[ht]
	\centering
    \includegraphics[width = 0.45\textwidth]{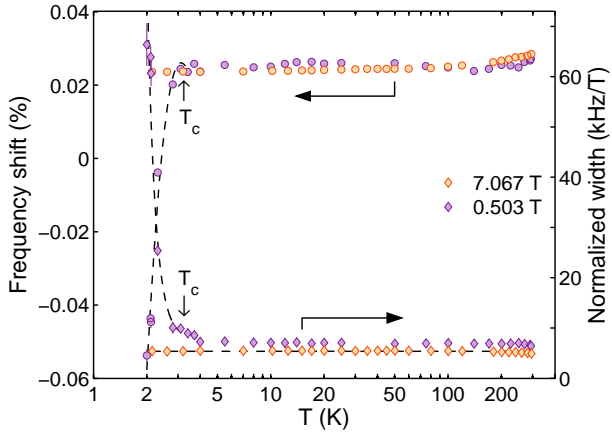}
	\vspace{-2ex}%
    \caption{Normalized $^{31}$P NMR Knight shifts $(\delta_\mathrm{iso}- \nu_0)/\nu_0$ 
    (left) and line widths 
    (right) vs.\  temperature, as measured at 0.5 and 7\,T. While there is 
    a clear drop of frequency and increase of width below $T_c$, the data 
    in the normal phase are practically temperature- and field-independent, 
    indicating an almost ideal metallic behavior.}
    \label{fig:NMR_width_shift}
\end{figure}

Typical $^{31}$P NMR lineshapes are shown in Fig.~\ref{fig:line_shapes} 
in the Appendix. As expected, the NMR lines  at 0.5\,T show a significant 
broadening and frequency shift below $T_c$ in the superconducting phase. 
The 7-T lines, instead, are practically temperature independent, with 
typical shapes reflecting the 
Knight shift anisotropy. The Knight shift, 
peak position, and linewidth were obtained using the \texttt{Dmfit} 
software \cite{Massiot2002}, 
with the best-fit parameters being $\delta_\mathrm{iso}=285$\,ppm, 
$\Omega=35$\,kHz, and $\eta_\mathrm{CS}=0.38$.
Here, $\delta_\mathrm{iso} = (\delta_{11} + \delta_{22} + \delta_{33})/3$ is 
the isotropic component of the shift tensor, i.e., the average of 
its diagonal components in the principal axes system (PAS), 
the linewidth is $\Omega = |\delta_{11}-\delta_{33}|$, and $\eta_\mathrm{CS} =
 (\delta_{22}-\delta_{11})/(\delta_{33}-\delta_\mathrm{iso})$.
The temperature evolution of the line widths and shifts 
(normalized to the field value and the reference frequency, respectively) 
is shown in Fig.~\ref{fig:NMR_width_shift}. 
The almost temperature- and field-independent line widths and 
frequencies, suggest a behavior close to that expected for 
weak- to moderately-correlated metal. The above picture changes only 
close to $T_c$, below which we detect a significant drop in frequency 
and increase of line width, reflecting the electron pairing and the 
development of the flux-line lattice in the superconducting phase, respectively.

%
\begin{figure}[ht]
	\centering
    \hspace*{-3.5mm}
    \includegraphics[width = 0.45\textwidth]{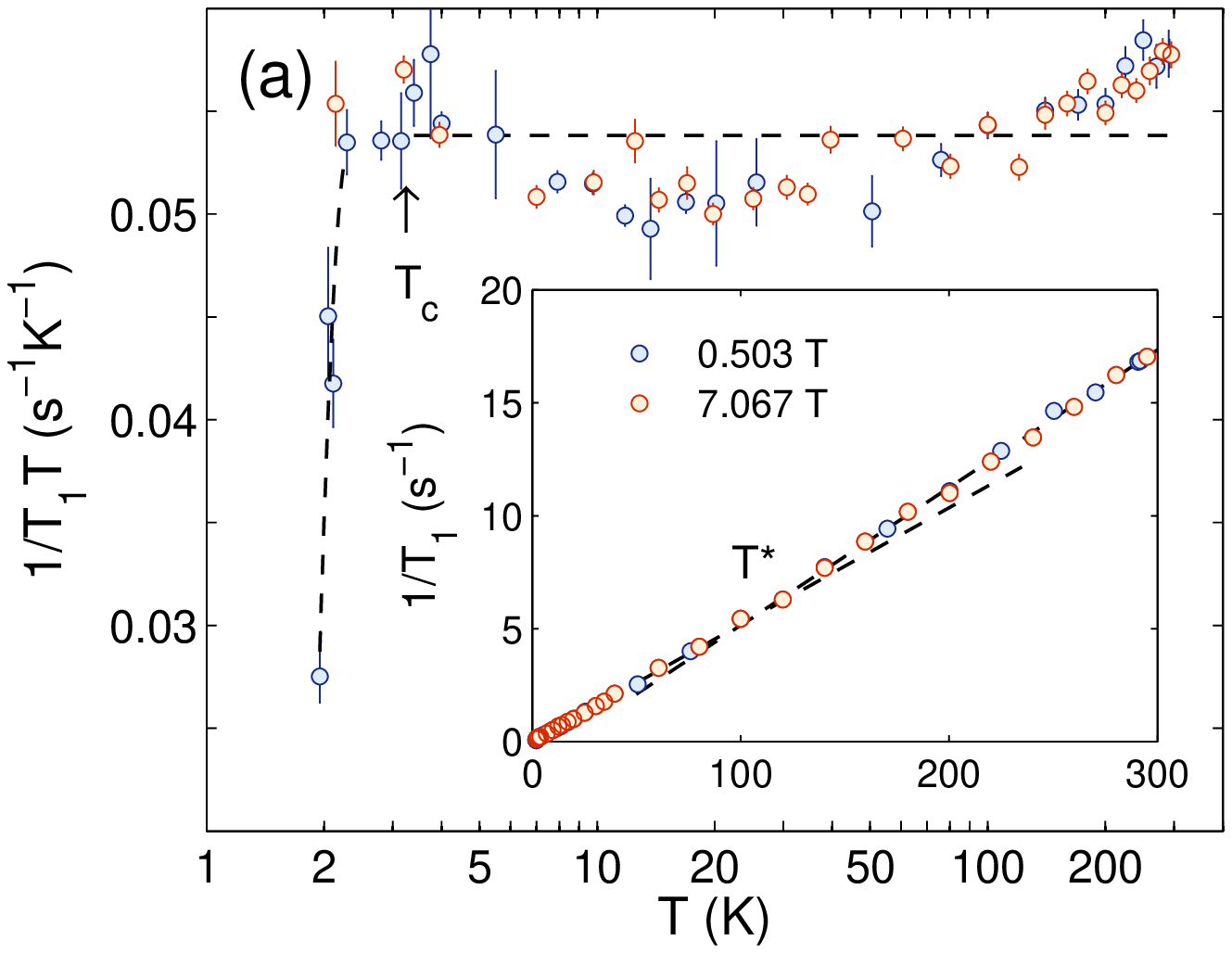}
    \includegraphics[width=0.44\textwidth,angle=0]{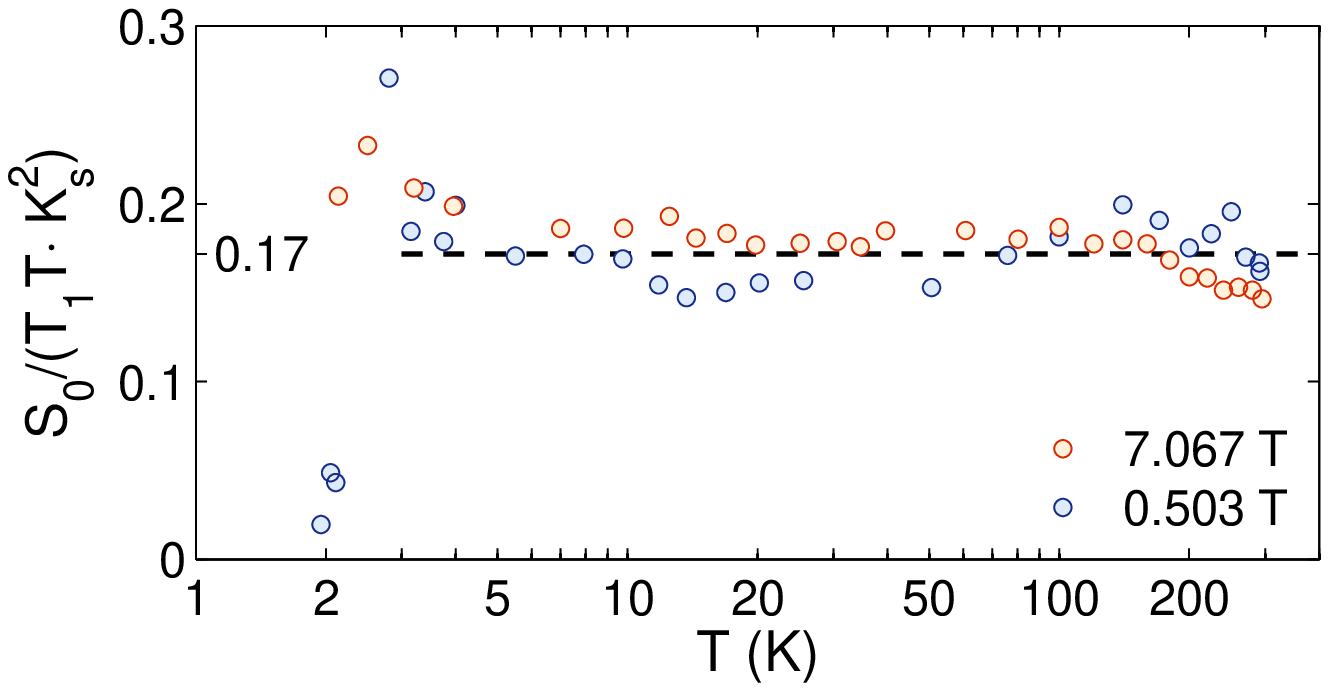}
\vspace{-2ex}%
   \caption{(a) $1/T_1T$ vs.\ temperature measured at 0.5- and 7\, T.
    Above $T_c$ the product is almost constant (0.054\,s$^{-1}$K$^{-1}$).
    Inset: $1/T_1$ vs.\ temperature measured at the same fields. The 
    field independent spin-lattice relaxation rate is linear in 
    temperature and changes slope at $T^\star \sim 103$\,K. 
    (b) The Korringa ratio vs.\ temperature is almost constant down to $T_{c}$. 
    An average $\alpha^{-1}$  value much lower than 1 
    (here shown by a dashed line) suggests 
    ferromagnetic electronic correlations in Mo$_{3}$P.
    \label{fig:T1_500mT_7T}}
\end{figure}

%
The temperature-dependent NMR relaxation rate $1/T_1$ provides useful 
insight into the dynamics of conduction electrons and their degree of 
correlation. The $1/T_1$ data, shown in Fig.~\ref{fig:T1_500mT_7T}(a) 
(inset), indicate an almost linear behavior with slightly different 
slopes in the regime above and below $T^\star \sim 103$\,K. While the
change in slope might reflect a change in the electron-nuclei 
coupling, the linearity of $1/T_1(T)$ is once more a clear indication 
of weak- to moderate electron correlations. 
%
%
This conclusion is confirmed by the $1/T_1T$ data shown in 
Fig.~\ref{fig:T1_500mT_7T}(a) (main panel), where the reported product, 
at a first approximation considered as constant, is 0.054\,s$^{-1}$K$^{-1}$.
Compared to other metallic superconductors, such as MgB$_2$, AlB$_2$, and 
ZrB$_2$ (see Fig.~5 in Ref.~\onlinecite{Barbero2017}), such a value indicates 
a good metallicity, i.e., a relatively high electron density of states at the Fermi level. 
Upon closer inspection we note that, after a minimum at $\sim 20$\,K, 
$1/T_1T$ increases slightly with temperature. The origin of this minimum 
is currently unknown to us.


The Korringa relation~\cite{Korringa1950} states the proportionality 
between the spin relaxation rate divided by the square of Knight shift 
(spin part) with the temperature. The total Knight shift 
$K$ consists of a spin- $K_{s}$ and an orbital $K_\mathrm{orb}$ part. 
Normally these can be separated by using the (typically linear) 
$K$--$\chi$ plot, with $K_\mathrm{orb}$ being the intercept value to 
the $K$ axis. However, as shown in Fig.~\ref{fig:CJ_plot_Mo3P}, 
Mo$_{3}$P exhibits a complex behavior, most likely reflecting the 
presence of $d$-type electrons (see Appendix). In our case, considering 
the opposite-spin pairing of electrons in the SC phase, we can assume 
$K_{s}$ to be fully suppressed at 0\,K. From Fig.~\ref{fig:NMR_width_shift} 
we can determine $K_\mathrm{orb} \sim -0.05$\% and, hence, $K_{s} \sim 0.07$\%.
Originally derived for simple $s$-band metals with negligible electronic 
correlations, $K_{s}$ was successively extended also to $d$-band 
metals, where core polarization effects typically dominate both the 
Knight shift and the relaxation rate~\cite{Yafet1964}.
The experimental value of the proportionality constant and its
temperature dependence, provide insight into the degree of electron 
correlations in the normal state of a material. In particular:
\begin{equation}
T_1TK_{s}^2 = \alpha S_0,\quad \text{with}\quad S_0 = \frac{\gamma_e^2}{\gamma_N^2}\frac{\hbar}{4\pi k_\mathrm{B}}, 
\end{equation}
where $\gamma_e$ is the gyromagnetic ratio for a free electron, and 
$\gamma_N$ the gyromagnetic ratio of the probe nucleus. Here, $\alpha$ 
is a coefficient of correlation which, in the ideal case of uncorrelated 
electrons, should be 1. 
As shown in Fig.~\ref{fig:T1_500mT_7T}(b), in 
Mo$_{3}$P the Korringa product is virtually temperature independent 
(above $T_c$), since in our case small variations in $1/T_1T$ are 
mostly compensated by those of Knight shift (see inset in 
Fig.~\ref{fig:CJ_plot_Mo3P}).
The resulting $\alpha$ value is 5.8, i.e., much higher than unity. 
Since the $^{31}$P NMR $K_s$ shift is mostly due to 4$d$-type conduction 
electrons (Mo), a high $\alpha$ value is not unusual (see Ref.~\cite{Narath1968}) 
and suggests at most moderate ferromagnetic correlations.



\subsection{\label{ssec:DFT} Electronic band structure and discussion}
%
\begin{figure}[ht]
	\centering
	\includegraphics[width = 0.48\textwidth]{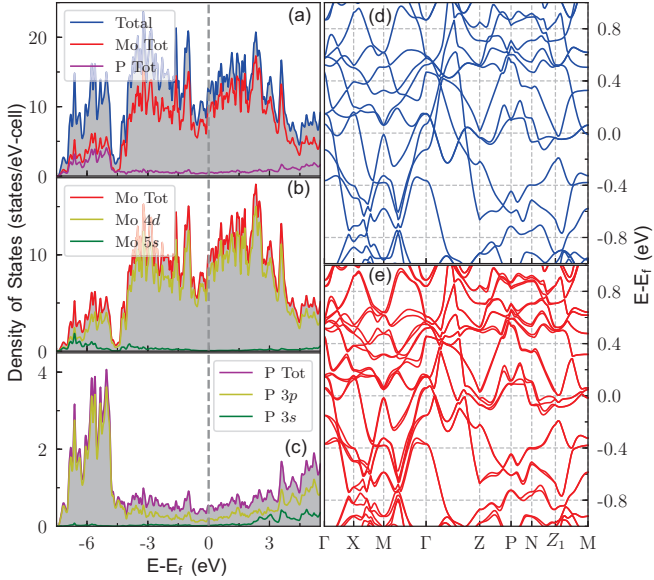}
	\caption{Calculated electronic band structure for Mo$_3$P. 
	Total and partial (Mo and P) density of states (a). 
	Orbital-resolved density of states for the Mo- (b) and P atoms (c). 
	Band structure of Mo$_3$P without- (d) and with SOC interactions (e), 
	calculated within $\pm$1\,eV from the Fermi energy level.}
	\label{fig:DOS}
\end{figure}
%
To further understand the properties of the normal- and superconducting 
states of Mo$_3$P, we performed DFT band-structure calculations, whose 
main results are summarized in Fig.~\ref{fig:DOS}. As shown in panels 
(a)-(c), close to Fermi level the DOS is dominated by the Mo 4$d$-electrons, 
while the low-lying states near $-6$\,eV are dominated by contributions 
from both Mo-$4d$ and P-$3p$ electrons. The estimated DOS at Fermi level 
is about 1.56 states/eV-f.u (= 12.5 states/eV-cell/$Z$, with $Z = 8$ 
the number of Mo$_3$P formula units per unit cell), which is 
comparable to the experimental results in Sec.~\ref{ssec:Cp_zero} and Table~\ref{tab:parameter}.
The relatively high DOS hints at a good metallic behavior in Mo$_3$P, confirmed by both electrical resistivity data and the fast NMR relaxation rates.
The electronic band structure 
calculated without considering the SOC is shown in panel (d), while that 
including SOC in panel (e). In the former case, there are two almost 
overlapping bands at the $Z$ point, about $\sim$0.05\,eV below the Fermi 
level. After including the SOC the bands separate, since SOC breaks the 
band degeneracy and brings one of the bands closer to the Fermi level 
[see Fig.~\ref{fig:DOS}(e)]. The dispersion of the Mo 4$d$ bands, which 
cross the Fermi level, is rather small with a bandwidth of 
about $\sim 0.3-0.4$\,eV.
The band splitting due to the ASOC is 
about $\sim$90\,meV, a value comparable to that of other NCSCs, e.g., 
Y$_2$C$_3$ ($\sim$15\,meV)~\cite{Nishikayama2007} 
and (Sr,Ba)(Ni,Pd)Si$_3$ ($<$ 20\,meV)~\cite{Kneidinger2014},
but much 
smaller than that of CePt$_3$Si ($\sim$200\,meV)~\cite{Samokhin2004} 
and Li$_2$Pt$_3$B ($\sim$200\,meV)~\cite{Lee2005}. It is accepted that 
the band splitting at Fermi level is of primary importance for the 
superconducting properties~\cite{Bauer2012}. Indeed, because of the 
large ASOC-induced band splitting, Li$_2$Pt$_3$B shows nodal 
superconductivity~\cite{Yuan2006}, while the upper critical field in 
CePt$_3$Si exceeds the Pauli limit~\cite{Bauer2004}. 
Conversely, in  Y$_2$C$_3$ and (Sr,Ba)(Ni,Pd)Si$_3$, the upper-critical-field is 
relatively small and they are fully-gapped 
superconductors~\cite{Kneidinger2014, Kuroiwa2008}. 
The latter seems also to be  the case of Mo$_3$P, which shows 
a small band splitting, a low $H_{c2}$ value, and a singlet-pairing 
superconducting state.

\begin{figure}[ht]
	\centering
	\includegraphics[width = 0.48\textwidth]{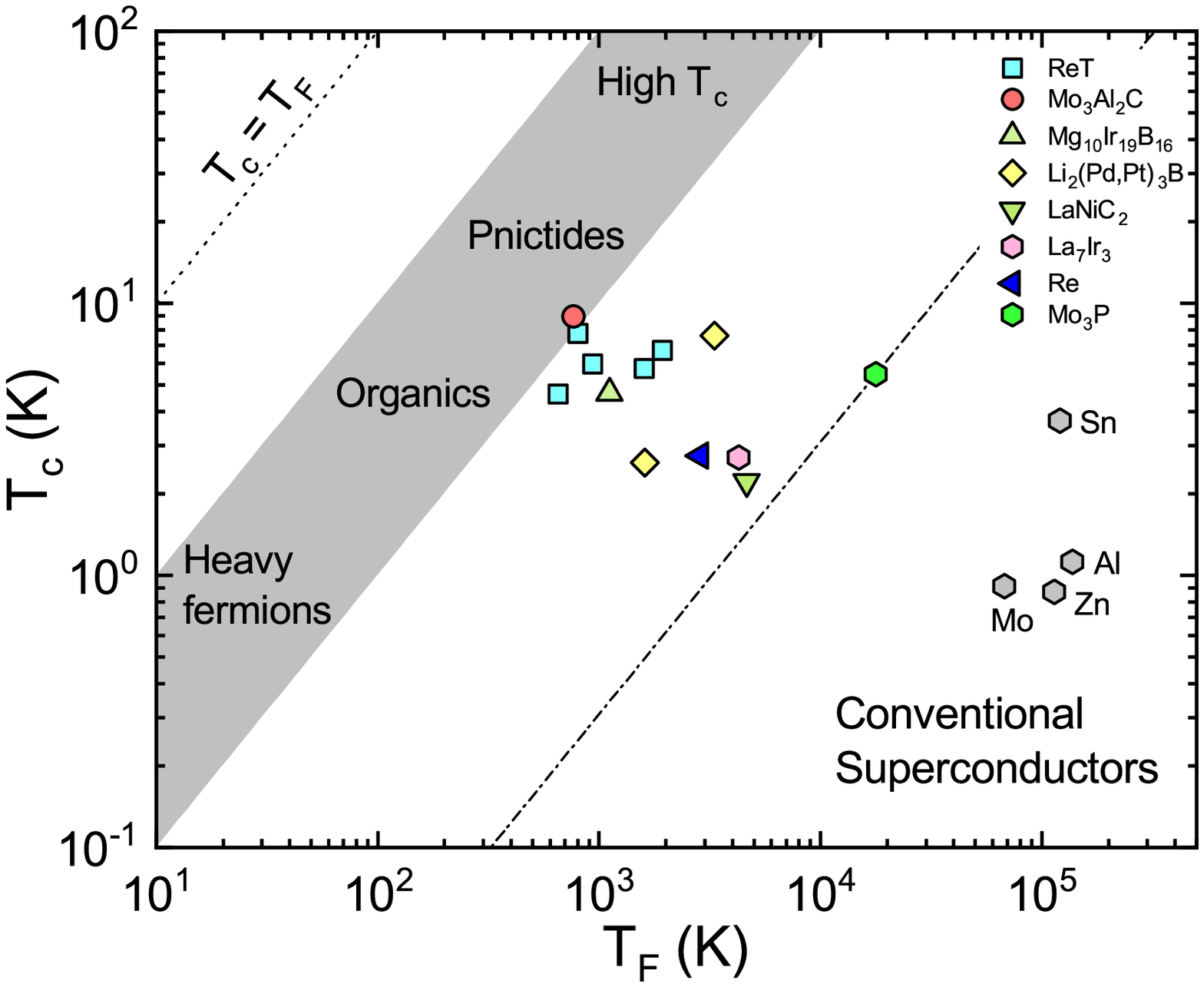}
	\caption{\label{fig:uemura} Uemura plot for different types of superconductors. 
	With  $1/100 < T_c/T_\mathrm{F}<1/10$, the grey region indicates the band of 
	unconventional superconductors, including heavy fermions, organics, fullerenes, 
	pnictides, and high-$T_c$ cuprates. Selected non\-cen\-tro\-sym\-met\-ric 
	and elementary superconductors are shown as symbols (adopted from 
	Refs.~\onlinecite{Barker2018,Uemura1991,TianReNb2018}). The dotted line  
	corresponds to $T_c = T_\mathrm{F}$ (here $T_\mathrm{F}$ is the 
	temperature associated with the Fermi energy, $E_\mathrm{F} = k_\mathrm{B} T_\mathrm{F}$), 
	while the dash-dotted line indicates $T_c/T_\mathrm{F} = 3.09 \times 10^{-4}$ for Mo$_3$P.}
\end{figure}

We also compare the superconductivity of Mo$_3$P with other NCSCs. As shown in Fig.~\ref{fig:uemura}, the different families of superconductors can be classified according to the ratio of the superconducting transition temperature $T_c$ to the effective Fermi temperature	$T_\mathrm{F}$, in a so-called Uemura plot~\cite{Uemura1991}. Several types of unconventional superconductors, including 	heavy-fermions, organics, high-$T_c$ iron pnictides, and cuprates, 
all lie in a $1/100 < T_c/T_\mathrm{F} <1/10$ range (grey region). Conversely, conventional BCS superconductors exhibit 
$T_c/T_\mathrm{F} <1/1000$, here exemplified by the elemental Sn, Al, Zn, and Mo. Most of the NCSCs, e.g., the Re$T$, Mo$_3$Al$_2$C, Li$_2$(Pd,Pt)$_3$B, and LaNiC$_2$, exhibit a $T_c/T_\mathrm{F}$ between the unconventional- and conventional case. This is also the case for Mo$_3$P. According to the superconducting parameters obtained from our measurements (see Table.~\ref{tab:parameter}), we find a $T_c/T_\mathrm{F}$ =  $ 3.09 \times 10^{-4}$ for Mo$_3$P, which is significantly enhanced  ($\sim$ 30 times) compared to elementary Mo ($T_c/T_\mathrm{F} = 0.92/6.77\times 10^4 = 1.36 \times 10^{-5}$)~\cite{Kittel2005}. On the other hand, Mo$_3$P shows a similar $T_c/T_\mathrm{F}$ value as LaNiC$_2$ and La$_7$Ir$_3$ (all lying close to the same dash-dotted line). While the latter exhibit unconventional superconductivity with broken TRS~\cite{Hillier2009, Barker2015}, the TRS is preserved in Mo$_3$P. Further studies, as e.g., gap-symmetry analysis, are desirable to offer an answer to such different outcomes.

\begin{table}[!bht]
	\centering
	\caption{Normal- and superconducting state properties of Mo$_3$P, as 
	determined from electrical resistivity, magnetic susceptibility, 
	specific-heat, NMR, and $\mu$SR measurements. The London penetration 
	depth $\lambda_\mathrm{L}$, the effective mass $m^{\star}$, bare 
	band-structure effective mass $m^{\star}_\mathrm{band}$, carrier 
	density $n_\mathrm{s}$, BCS coherence length $\xi_0$, electronic 
	mean-free path $l_e$, Fermi velocity $v_F$, and effective Fermi 
	temperature $T_F$ were estimated following the Eqs.\,(40)--(50) 
	in Ref.~\onlinecite{Barker2018}.\label{tab:parameter}}
	\begin{ruledtabular}
		\begin{tabular}{lcl}
			Property                               & Unit            & Value     \\ \hline
			$T_c$\footnotemark[1]                  & K               & 5.5(2)    \rule{0pt}{2.6ex} \\
			$\rho_0$                               & $\mu\Omega$cm   & 5.92(5)   \\
			$\Theta_\mathrm{D}^\mathrm{R}$         & K               & 251(2)    \\[2mm]
			$\mu_0H_{c1}$                          & mT              & 29.4(2)   \\
		    $\mu_0H_{c1}$$^{\mu\mathrm{SR}}$       & mT              & 24.9(5)   \\
		    $\mu_0H_{c2}$$^\rho$                   & T               & 1.09(1)   \\
			$\mu_0H_{c2}$$^{\chi,C}$             & T               & 0.94(1)   \\
		    $\mu_0H_{c2}$$^{\mu\mathrm{SR}}$       & T               & 0.96(3)   \\[2mm]
 	 	    $\xi(0)$                               & nm              & 18.7(1)   \\
 	 	    $\gamma_n$                             & mJ/mol-K$^2$    & 10.3(4)   \\
 	 	    $\Theta_\mathrm{D}^\mathrm{C}$         & K               & 225(3)    \\
 	 	    $\lambda_\mathrm{ep}$                  & ---             & 0.72(1)   \\[2mm]
 	 	    $N(\epsilon_\mathrm{F})$               & states/eV-f.u.  & 2.2(1)    \\
 	 	    $N_\mathrm{band}(\epsilon_\mathrm{F})$ & states/eV-f.u.  & 1.28(1)   \\
 	 	    $N(\epsilon_\mathrm{F})^\mathrm{DFT}$  & states/eV-f.u.  & 1.56      \\
 	 	    $E_\mathrm{ASOC}$                      & meV             & 90        \\[2mm]
 	 	    $\Delta_0$$^{\mu\mathrm{SR}}$          & meV             & 0.83(1)   \\  
 	 	    $\Delta_0^{C}$                        & meV             & 0.82(1)   \\
 	 	    $\Delta C/\gamma_\mathrm{n}T_c$        & ---             & 1.44(1)   \\
 	 	    $\lambda_0$\footnotemark[2]            & nm              & 126(1)    \\
 	 	    $\lambda_0$                            & nm              & 121(2)    \\
 	 	    $\lambda_\mathrm{GL}$                  & nm              & 113(1)    \\
 	 	    $\lambda_\mathrm{L}$                   & nm              & 63(1)     \\[2mm]
 	 	    $m^{\star}$                            & $m_e$           & 5.4(2)    \\
 	 	    $m_\mathrm{band}$                      & $m_e$           & 3.1(2)   \\[2mm]
 	 	    $n_\mathrm{s}$                         & 10$^{28}$\,m$^{-3}$ & 3.86(3) \\
 	 	    $\xi_0$                                & nm              & 56 (1)    \\
 	 	    $l_e$                                  & nm              & 18.8(2)    \\
 	 	    $v_\mathrm{F}$                         & 10$^5$\,ms$^{-1}$ & 2.22(7)  \\
 	 	    $T_\mathrm{F}$                         &  10$^4$\,K      & 1.78(6)    \\
		\end{tabular}
		\footnotetext[1]{Similar values were determined via electrical resistivity, magnetic sus\-cep\-ti\-bi\-li\-ty, and heat-capacity measurements.}.
		\footnotetext[2]{Derived from a fit to Eq.~\eqref{eq:TF_muSR_H}.}	
	\end{ruledtabular}
\end{table}

\section{\label{ssec:Sum}Conclusion}
To summarize, we studied the normal- and superconducting state properties 
of the Mo$_3$P NCSC by means of bulk- (electrical resistivity, magnetization, 
and heat capacity), local-probe ($\mu$SR and NMR) techniques, and numerical 
band-structure calculations. The superconducting state of Mo$_3$P is 
characterized by $T_c = 5.5$\,K, a relatively low $H_{c2}(0) = 0.94$\,T, 
and $\kappa \sim 6$. The temperature dependence of the superfluid density 
and zero-field electronic specific heat reveal a \emph{nodeless} 
superconducting state, which is well described by an 
\emph{isotropic $s$-wave} model and is consistent with \emph{spin-singlet 
pairing}. The lack of spontaneous magnetic fields below $T_c$ indicates 
that time-reversal symmetry is \emph{preserved} in the superconducting state of 
Mo$_3$P. Electronic band-structure calculations suggest a small band 
splitting due the ASOC and a sizable density of states at Fermi level, 
confirmed also by NMR, which shows Mo$_3$P to be a metal with a rather 
high $N(\epsilon_\mathrm{F})$ and moderate ferromagnetic 
electron correlations.

\begin{acknowledgments} 
This work was supported by the Schwei\-ze\-rische Na\-ti\-o\-nal\-fonds zur 
F\"{o}r\-de\-rung der Wis\-sen\-schaft\-lich\-en For\-schung, SNF 
(Grants no.\ 200021-169455 and 206021-139082). L.\ J.\ C.\ thanks the 
MOST Funding for the support under the projects 104-2112-M-006-010-MY3 and 
107-2112-M-006-020. 
 We also acknowledge the assistance from other beamline scientists on HAL-9500 and GPS $\mu$SR spectrometers at PSI.
\end{acknowledgments}

\appendix*
\section{NMR line shapes and relaxation rates}

\begin{figure*}
\centering
\includegraphics[width=0.322\textwidth]{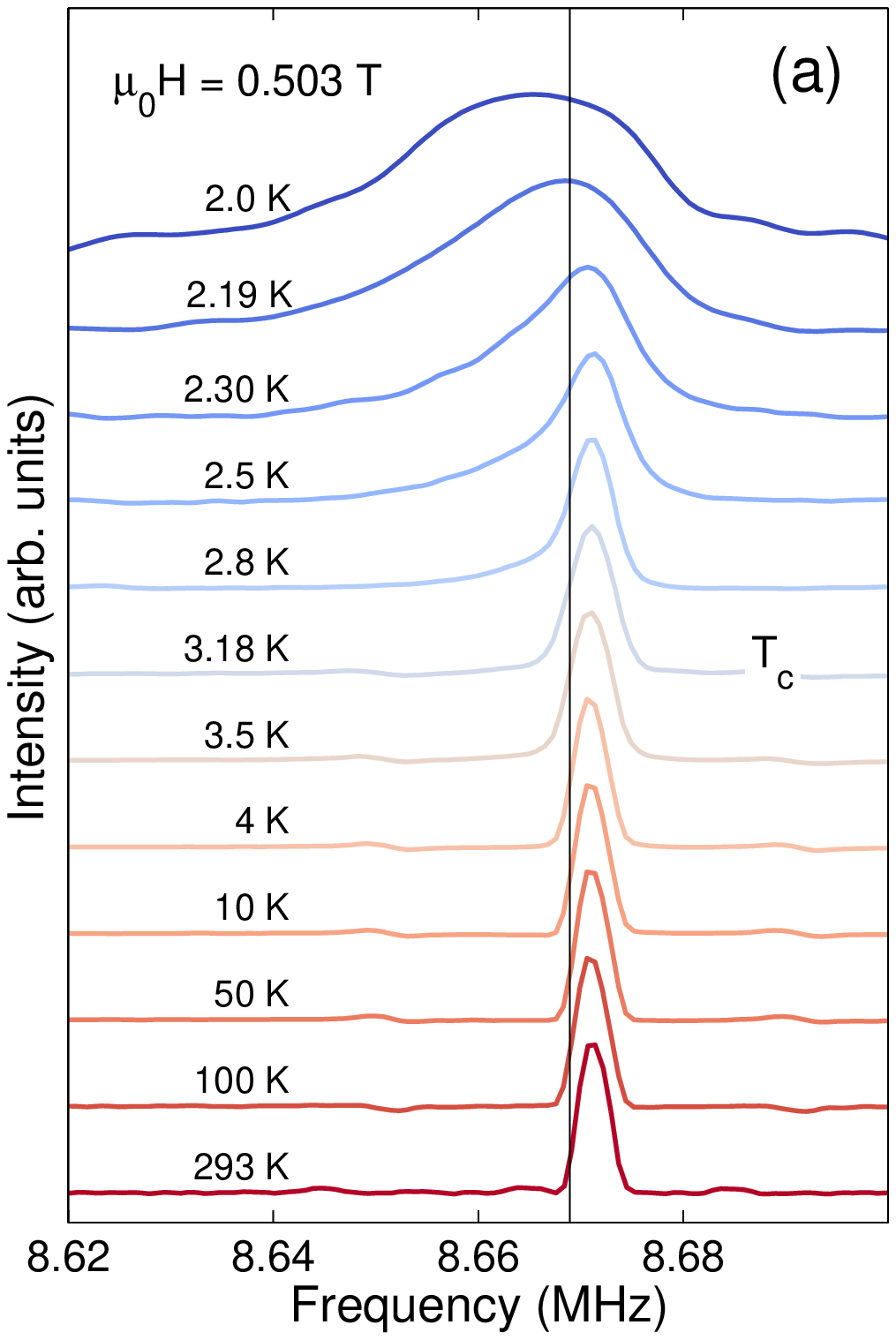}
\hspace{5mm}
\includegraphics[width=0.33\textwidth]{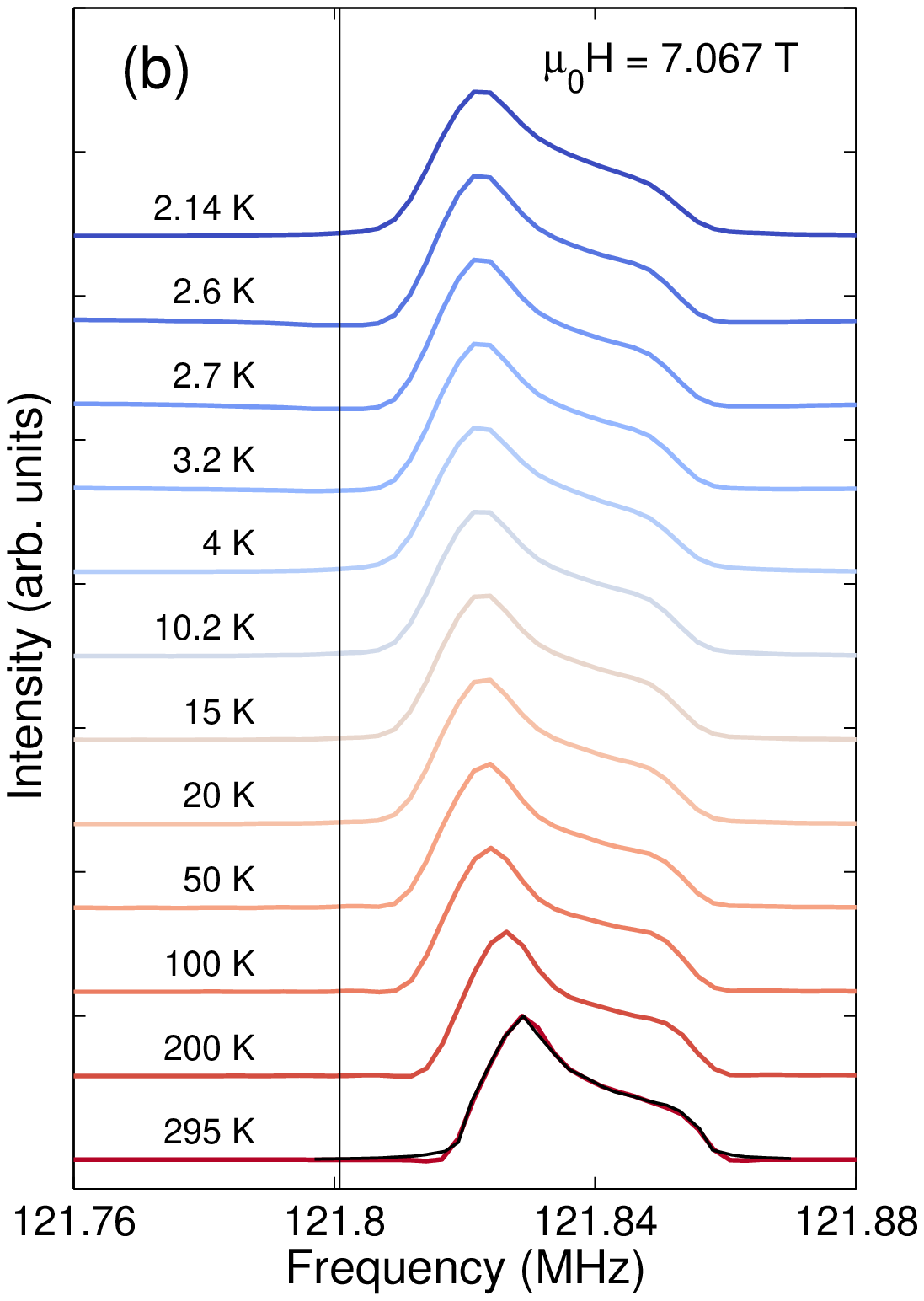}
\caption{Selected $^{31}$P NMR line shapes in Mo$_{3}$P measured at 
(a) 0.5- and (b) 7\,T. While at low field a temperature-dependent 
line broadening and frequency shift are observable (especially below 
$T_{c}$), at high field the lineshapes are practically temperature-independent. 
The vertical lines indicate the $^{31}$P NMR reference frequencies at 
the respective fields. The black line on top of the 295\,K data in 
panel (b) represents a fit using a Knight-shift anisotropy model 
(see text for details).\label{fig:line_shapes}}
\end{figure*}

A selection of $^{31}$P NMR lineshapes for both magnetic fields (0.5 
and 7\,T)  is shown in Fig.~\ref{fig:line_shapes}. At low field (a), 
the transition to the superconducting phase at low temperature is 
indicated by a broadening of the line and by its qualitative change 
of shape. Above $T_c$, the line shape is described by a Gaussian 
(analysis performed by \texttt{Dmfit}), whereas below $T_c$, the 
lineshapes can be fitted by means of a convolution of a Gaussian with a 
Lorentzian. At high field (b), the SC phase transition is suppressed 
by the applied magnetic field, which also enhances the anisotropy of 
the lineshapes. The black line on top of the 295-K data represents a 
fit using a Knight-shift anisotropy model (see main text). 
Here, $\delta_{11} > \delta_{22} > \delta_{33}$ represent the principal 
components of the shift tensor, identified with the shoulders and the 
maximum  ($\delta_{22}$) of the NMR lineshapes shown in 
Fig.~\ref{fig:line_shapes}(b).
The fit parameters in the Herzfeld-Berger notation 
\cite{Herzfeld1980} used here are the isotropic component of the 
shift tensor $\delta_\mathrm{iso}$ (i.e., the trace average), 
the linewidth $\Omega$, and the anisotropy $\eta_\mathrm{CS}$ (see main text).

\begin{figure}
\centering
\includegraphics[width=0.4\textwidth]{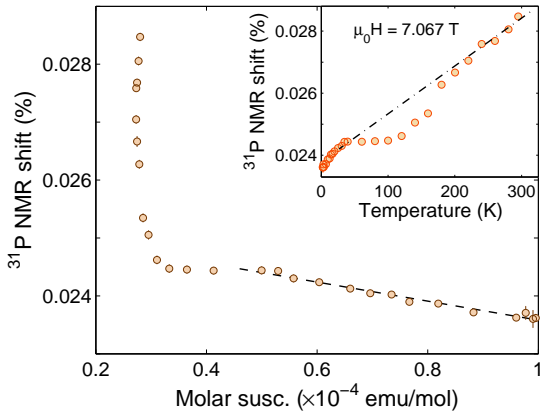}
\caption{Clogston-Jaccarino $K$--$\chi$ plot for Mo$_3$P 
as measured at 7.067\,T. The low-temperature part was fitted by 
using a linear relation, as given in Eq.~\eqref{eq:CJ_fit}. Inset: 
$^{31}$P NMR shift vs.\ temperature measured at the same field. 
The dash-dotted line highlights the shift deviations in the 
50--200\,K range.\label{fig:CJ_plot_Mo3P}}
\end{figure}

The Clogston-Jaccarino $K$--$\chi$ plot shown in Fig.~\ref{fig:CJ_plot_Mo3P} 
clearly indicates that the Knight shift depends linearly on the molar 
susceptibility $\chi$ in the range 0.5--$1\times 10^{-4}$\, emu$/$mol, 
corresponding to a temperature range 2--50\,K. This linear dependency 
is given by: 
\begin{equation}
\label{eq:CJ_fit}
K(\chi) [\%] =  -1.6451 \times 10^{-3}\chi + 2.5226 \times 10^{-2}.
\end{equation}
The offset may indicate the presence of other temperature-independent 
components. The coupling constant [i.e., the coefficient in 
Eq.~\eqref{eq:CJ_fit}] may become temperature-dependent for 
$\chi < 0.5\times 10^{-4}$\,emu$/$mol \cite{Ho2014}. From the slope 
$\mathrm{d} K /\mathrm{d} \chi$ of the $K$--$\chi$ plot for $T < 50$\,K, 
by assuming a standard electronic $g$-factor of 2.0, the relation 
$K = gA_\mathrm{hf}\chi + K_\mathrm{offset}$ gives a hyperfine coupling 
constant $A_\mathrm{hf} = -0.082$\,T$/\mu_\mathrm{B}$ and a zero 
intercept of 0.025\%.

As shown in Fig.~\ref{fig:CJ_plot_Mo3P}, both the $K$--$\chi$ plot behavior 
and the temperature dependence of the NMR shift are rather complicated. This 
is not unusual for $d$-type electron systems, known for their 
nonlinear $K(T)$ dependencies (see, e.g., Ref.~\cite{Carter1977}).
In our case, between approximately 30 and 200\,K, the Knight shift deviates 
markedly from linearity to exhibit a shallow minimum around 100\,K. 
At 30\,K, $1/(T_1T)$ exhibits a minimum [see Fig.~\ref{fig:T1_500mT_7T}(a)], 
while around 100\,K, $1/T_1$ 
changes slope. Since these anomalies appear in rather independent 
types of measurements ($K$ and $T_1$), they might represent true 
modifications of the Mo$_3$P electronic properties. Although, currently an 
explanation concerning their nature is missing, future studies of the Hall 
coefficient vs.\ temperature might help to better determine their origin.

\bibliography{Mo3P_bib}

\end{document}